\documentclass[iop]{emulateapj}

\usepackage{amsmath,amssymb,latexsym,graphics,epsfig,exscale,graphicx}
\usepackage{pstricks}
\usepackage{multirow}
\usepackage{graphicx}
\usepackage{subfigure}
 
\newpsobject{showgrid}{psgrid}{subgriddiv=1}

\begin{document}

\title{Distribution of Maximal Luminosity of Galaxies in the Sloan Digital Sky Survey}
\shorttitle{Distribution of Maximal Luminosity of Galaxies in the Sloan Digital Sky Survey}
\shortauthors{Taghizadeh-Popp et al.}
\author{M. Taghizadeh-Popp \altaffilmark{1}, K. Ozog\'{a}ny \altaffilmark{2}, Z.  R\'{a}cz \altaffilmark{2}, E. Regoes \altaffilmark{3}, A.S. Szalay \altaffilmark{1}}
\altaffiltext{1}{Department of Physics and Astronomy, Johns Hopkins University. 3400 North Charles Street, Baltimore, MD 21218, USA. e-mail: mtaghiza [at] pha.jhu.edu}
\altaffiltext{2}{Institute for Theoretical Physics - HAS,  E\"{o}tv\"{o}s University, P\'{a}zm\'{a}ny  s\'{e}t\'{a}ny 1/a, 1117 Budapest, Hungary}
\altaffiltext{3}{European Laboratory for Particle Physics (CERN), Geneva, Switzerland}

%\affiliation{Institutes}
\date{\today}

\begin{abstract}

Extreme value statistics (EVS) is applied to the distribution of galaxy luminosities in the Sloan Digital Sky Survey (SDSS). We analyze the DR8 Main Galaxy Sample (MGS), as well as the Luminous Red Galaxies (LRG). Maximal luminosities are sampled from batches consisting of elongated pencil beams in the radial direction of sight. For the MGS, results suggest a small and positive tail index $\xi$, effectively ruling out the possibility of having a finite maximum cutoff luminosity, and implying that the luminosity distribution function may decay as a power law at the high luminosity end. Assuming, however, $\xi=0$, a non-parametric comparison of the maximal luminosities with the Fisher-Tippett-Gumbel distribution (limit distribution for variables distributed by the Schechter fit) indicates a good agreement provided uncertainties arising both from the finite batch size and from the batch size distribution are accounted for. For a volume limited sample of LRGs, results show that they can be described as being the extremes of a luminosity distribution with an exponentially decaying tail, provided the uncertainties related to batch-size distribution
are taken care of.

\end{abstract}


%%%%%%%%%%%%%%%%%%%%%%%%%%%%%%%%%%%%%%%%%%%%%%%%%%%%%%%%%%%%%%%%%%%%%%%%%%%%%%%%%%%%%%%%%%%%%%%%%%%%%%%%%%%%%%%%%%%%%%%%%%%%%%%%%%%%%%%%%%%%%%%%%%%%%%%%%%%%%%%%%%%%%%%%%%%%%%%%%%%%
%%%%%%%%%%%%%%%%%%%%%%%%%%%%%%%%%%%%%%%%%%%%%%%%%%%%%%%%%%%%%%%%%%%%%%%%%%%%%%%%%%%%%%%%%%%%%%%%%%%%%%%%%%%%%%%%%%%%%%%%%%%%%%%%%%%%%%%%%%%%%%%%%%%%%%%%%%%%%%%%%%%%%%%%%%%%%%%%%%%%

\section{Introduction}

Extreme value statistics is a powerful tool for analyzing the behavior of the tails of distributions. It is well-known that the distribution of extreme values for a sample of $N$--i.i.d. (independent, identically distributed) random variables converge (as $N\rightarrow \infty$) to a few limiting distributions depending on the tail behavior of the parent population, namely Fisher-Tippett-Gumbel, Weibull and Fisher-Tippett-Frechet \citep{gumbel,galambos,embrechts1997,reiss1997,coles2001}. However, the onset of this finite sample size scaling behavior is quite slow, and therefore requires very large samples to converge. This is the primary reason why astronomy has seen few applications of EVS to date.

The emergence of dedicated wide angle galaxy surveys, such as SDSS \citep{stoughton2002}, has made possible an increase in statistics, making galaxy samples in the SDSS redshift survey just large enough to attempt an analysis of the finite sample size scaling for all galaxies.
Here we chose to study the distribution of maximal luminosities of galaxies, since the galaxy luminosity distribution per volume or luminosity function (LF) is one of the most basic statistic measured in galaxy surveys. This function has been well described by a gamma distribution or so-called Schechter function \citep{schechter1976}, functionally similar (and motivated by) the theoretically derived Press-Schechter formula \citep{pressschechter1974}, with a power law distribution at the faint end and an exponentially falling tail at the bright end. When galaxies are grouped according to their morphologies, their respective LFs seem to belong to different classes, including bell-shaped distributions as well as gamma functions of different shape and scale parameters \citep{binggeli1988}. Current modeling of the conditional LF (CLF) to galaxy clusters in dark matter halos of a certain mass include the presence of central or brightest cluster galaxies (BCGs) with a log-normal CLF, while the rest of the galaxies (satellites) are given a power law CLF with a finite cut at high luminosities \citep{cooray2005,cooray2006}.

Special attention has always been paid to the high luminosity tail of the all galaxy LF. BCGs are the brightest of the old populations of red elliptical galaxies found in the high density cores of galaxy clusters are thought to have their progenitors formed at high redshift ($z \gtrsim 3$), and then have undergone a set of dry mergers in their life history \citep[e.g.][]{ostriker1977,delucia2007}. Their importance lies in the low scatter of their luminosities, making them useful as standard candles \citep{postman1995,loh2006,lin2010,paranjape2011,dobos2011}.

Several studies have been made to elucidate whether BCGs are the extremes of a red or early type galaxies LF or they come from other luminosity distribution. In order to answer the question, \cite{geller1976}, \cite{tremaine1977} and \cite{bhavsar1985} investigated the statistics derived from the first and second brightest luminosities (and the gap between them) in galaxy clusters. Their results based on smaller samples has been confirmed by \cite{loh2006}, who found that the luminosity gap between first and second-ranked galaxies is substantially larger than what can be explained with an exponentially decaying luminosity function. On the other hand, \cite{lin2010} shuffled the data to combine all galaxies of clusters to form a composite cluster, finding that BCGs in high luminosity clusters are not drawn from the luminosity distribution of all red cluster galaxies, while BCGs in less luminous clusters are consistent to be the statistical extreme. 

These previous studies were mainly directed toward the luminosity statistics within galaxy clusters. In this paper we will study galaxy luminosities as a whole, and the sampling will not be restricted to the maximal luminosities from galaxy clusters. This will keep the sample size large enough for studying the finite-size scaling behavior.

Since EVS is well known only for i.i.d. variables, one approach we will follow is trying to minimize the correlations between luminosities and positions by selecting the maximal luminosities from batches or blocks of galaxies in elongated regions or pencil beams along the line of sight, and defined by the footprint of the HEALPix tessellation on the sky \citep{gorski2005}. As we shall discuss, such elongated cells combined with the short range correlations in luminosities make possible an analysis of EVS based on the assumption that the luminosities approximate well an i.i.d. behavior. 
This approach allows us to show a working example designed to mimic the standard block maxima sampling method from EVS of time series, but also generalizing it to the case of variable block size, as discussed later in this paper.
Another simpler approach we will use for comparing with the previous method is the random sampling of the luminosity parent distribution in batches of fixed size.
These are new approaches for testing the bright end of the overall LF, and inherently different from previous studies that considered testing the luminosity extremes in galaxy clusters.

Within the i.i.d. framework, the shape of the galaxy luminosity function is important for the EVS. The exponential tail in the high luminosity end of the LF would imply a Fisher-Tippett-Gumbel (FTG) EVS distribution, with corrections for the finite sample sizes depending on the power law at lower luminosities \citep{Gyorgyi4}. In this analysis we will test the agreement with these expectations, and the analysis will also reveal whether or not a sharp cutoff at a high but finite luminosity exists.

We emphasize that even though the SDSS sample is large, the residual from the FTG distribution can be explained only when we consider the corrections due to both the finite size of the samples of each HEALPix pencil beam and the distribution present in the sample sizes (the number of galaxies in a cone is finite and varies from cone to cone). Thus we have here a pioneering example where a generalized finite size scaling (including sample-size distribution) is relevant in the data analysis.

The arguments and results will be presented in the following order. In Section \ref{Sec:SampleCreation} we describe our galaxy sample. Section \ref{LumFun} shows the fits to the galaxy luminosity distributions and functions. Section \ref{section:FootPrintCreation} explains the construction of the pencil beams and distribution of galaxy counts inside them. Section \ref{section:TheoryOfEVS} contains a discussion of the basic concepts of extreme value statistics with emphasis on possible deviations from the expected limit distributions due to finite number of the galaxies in the pencil beams and, furthermore, due to the pencil-to-pencil fluctuations in the galaxy counts. In Section \ref{Section:DistributionOfMaximalLuminosities} we present the results about the distribution of maximal luminosities with the conclusion that within the uncertainties coming from the finiteness of samples and from the sample--size distribution, the Fisher--Tippett--Gumbel distribution gives an excellent fit. The final remarks and discussion can be found in Section \ref{disc}.

Along this paper, we use the ($\Omega_{\rm L}$, $\Omega_{\rm M}$, $h_{0}$, $w_{0}$) = (0.7, 0.3, 0.7, -1) cosmology.

%%%%%%%%%%%%%%%%%%%%%%%%%%%%%%%%%%%%%%%%%%%%%%%%%%%%%%%%%%%%%%%%%%%%%%%%%%%%%%%%%%%%%%%%%%%%%%%%%%%%%%%%%%%%%%%%%%%%%%%%%%%%%%%%%%%%%%%%%%%%%%%%%%%%%%%%%%%%%%%%%%%%%%%%%%%%%%%%%%%%
%%%%%%%%%%%%%%%%%%%%%%%%%%%%%%%%%%%%%%%%%%%%%%%%%%%%%%%%%%%%%%%%%%%%%%%%%%%%%%%%%%%%%%%%%%%%%%%%%%%%%%%%%%%%%%%%%%%%%%%%%%%%%%%%%%%%%%%%%%%%%%%%%%%%%%%%%%%%%%%%%%%%%%%%%%%%%%%%%%%%

\section{Sample Creation}\label{Sec:SampleCreation}

In this paper we use photometric and spectroscopic data of galaxies from SDSS-DR8 \citep{york00,stoughton2002,aihara2011}, available in a MS-SQL Server 
database that can be queried online via CasJobs \footnote{ {\tt http://casjobs.sdss.org} }, and analyzed directly inside the database using an integrated cosmological functions library \citep{taghizadeh-popp2010}.
The galaxies studied were the DR7 legacy spectroscopically-targeted Main Galaxy Sample (MGS) \citep{strauss2002}, as well as the luminous red galaxies (LRGs) \citep{eisenstein2001}. The sky footprint of the clean spectroscopic survey builds up from a complicated geometry defined by {\tt sectors}, which cover a fractional area $F_{A}\simeq 0.1923$ of the whole sky. Redshift incompleteness arises from the fact that two 3'' aperture spectroscopic fibers cannot be put together closer than 55'' in the same plate. As a strategy, denser region in the sky are given a greater number of overlapping plates. However, only $\sim$93$\%$ (MGS) and $\sim$95$\%$ (LRG)  of the initial galaxies photometrically targeted have their spectra taken.

Several selection cuts and flags were applied in order to have a clean sample. We selected only {\tt science primary} objects classified as galaxies and appearing in calibrated images having the {\tt photometric} status flag. We used the {\tt score} quantity as a measure of the field quality with respect to the sky flux and the width of the point spread function, and selected only the fields in the range $0.6 \leq {\tt score} \leq 1.0$.
Furthermore, we neglected individual objects with bad deblending flags ({\tt PEAKCENTER, DEBLEND$\_$NOPEAK, NOTCHECKED  }) and interpolation problems ({\tt PSF$\_$FLUX$\_$INTERP, BAD$\_$COUNTS$\_$ERROR}) or suspicious detections ({\tt SATURATED NOPROFILE }), as well with problems in the spectrum ({\tt ZWARNING}) \footnote{Detailed explanation in {\tt sdss3.org/dr8/algorithms}}. 

\begin{figure}[h]
\epsscale{1.0}
\plotone{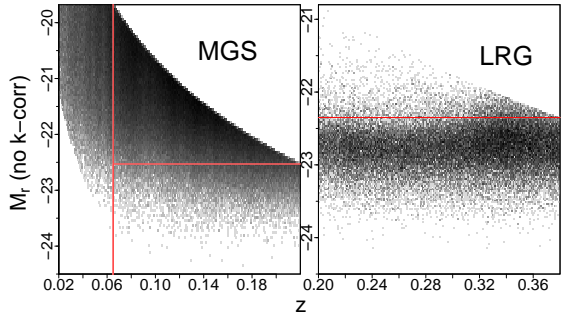}
\caption{Completeness limits for the raw galaxy samples. Plotted are the petrosian absolute Magnitude v/s redshift histogram (log-scaled) for the full MGS and LRG samples. No k-correction nor evolution is applied at this point. The red horizontal lines show the absolute magnitude limits for a complete sample (-22.53 and -22.35 for MGS and LRG respectively). Note that for redshifts greater than z=0.065 (vertical red line), the MGS becomes complete at the bright end. This redshift limit was also checked using the Vmax method explained in Sec. \ref{LumFun}.}
\label{Fig:AbsMagVSredshift}
\end{figure}

With respect to the MGS, they were observed as a magnitude limited sample, with a targeted r-band petrosian apparent magnitude cut of $m_r \leq 17.77$, and a redshift distribution peaking at $z \sim 0.1$. We further restrict this sample to safe cuts of $[m_{r,1},m_{r,2}]=[13.5,17.65]$. The lower limit is set due to the arising cross talk from close fibers in the spectrographs when they carry light from very bright galaxies, whereas the upper limit safely avoids the slight variations in the targeting algorithm of the limiting apparent magnitude around 17.77 over the sky. As shown in Fig. \ref{Fig:AbsMagVSredshift}, we chose galaxies in the redshift interval $[z_{1},z_{2}]=[0.065,0.22]$, since at redshift lower than $z_{1}$, the galaxy high luminosity tail becomes incomplete (due to imposing the apparent magnitude cut at $m_{r,1}$). This left us with $N_{g}=348975$ MGS galaxies in a volume of $V_{\rm S} = [V(z_{2})-V(z_{1})]\times F_{A} = 0.559 {\rm Gpc}^{3}$. 

With respect to the LRGs, they where selected from color cuts (in g-r v/s r-i space) in such a way that are traced across redshift as an old population of luminous and passively evolving red early type galaxies \citep{eisenstein2001}. This was done by modeling them with an old stellar population spectral template from PEGASE \citep{fioc1997}. We use LRGs in the CUT I sample, which was built to be almost a volume limited sample up to redshift 0.38 with an r-band petrosian apparent magnitude cut of $m_{r,2}=19.2$. We apply a safe redshift window of $[z_{1},z_{2}]=[0.20,0.38]$ since, at lower redshift, the color selection cuts admit blue galaxies belonging to the MGS. We further constraint this LRGs by considering galaxies whose r-band surface light profile can be modeled mainly as a DeVacouleurs profiles (as in elliptical galaxies) more than an exponential disc ({\tt fracDeV} $\geq 0.9$) \citep{strateva2001}. We also use the r-band concentration index R90/R50 \citep{shimazaku2001,strateva2001} to select mostly elliptical galaxies (R90/R50$\geq 2.7$ ). This left us with $N_{g}=$52579 LRGs in a volume of $V_{\rm S}=2.18 {\rm Gpc}^{3}$.

Since our samples span broad redshift and time intervals, it is crucial to apply a (k+evolution)-correction to $M_{r}$ in the form $M_{r}$ = $m_{r}$ - $DM(z)$ - $k(z)$ - $e(z)$, which brings all the galaxies to a common $z=0$ restframe. The k-corrections for the MGS were calculated by modeling each galaxy spectrum as the closest non negative linear combination of spectra drawn from the \cite{bruzual2003} templates (see \cite{budavari2000} and \cite{csabai2000}). We applied a simple average evolution correction as a linear function of redshift, derived by \cite{blanton2003} as 
 e(z)=-Qz (Q=1.62 for r-band, Q=4.22 for u-band). 
For the LRG case, we used the k+evolution correction derived from the PEGASE template. This was modeled as a 4th order polynomial in redshift, as used in \cite{loh2004} and \cite{loh2006}, where k(z)+e(z)=$0.115z$ +5.59$z^{2}$-24.0$z^{3}$+36.0$z^{4}$. 

We finally checked the first 1000 images of galaxies for each sample ranked by brightest r-band petrosian absolute magnitude, and rejected the objects whose photometry appears to be ruined by the leaked light of a nearby star. Also, objects where rejected in the case when the petrosian magnitude was more different than 0.8 magnitudes compared with the model magnitude.

%%%%%%%%%%%%%%%%%%%%%%%%%%%%%%%%%%%%%%%%%%%%%%%%%%%%%%%%%%%%%%%%%%%%%%%%%%%%%%%%%%%%%%%%%%%%%%%%%%%%%%%%%%%%%%%%%%%%%%%%%%%%%%%%%%%%%%%%%%%%%%%%%%%%%%%%%%%%%%%%%%%%%%%%%%%%%%%%%%%%
%%%%%%%%%%%%%%%%%%%%%%%%%%%%%%%%%%%%%%%%%%%%%%%%%%%%%%%%%%%%%%%%%%%%%%%%%%%%%%%%%%%%%%%%%%%%%%%%%%%%%%%%%%%%%%%%%%%%%%%%%%%%%%%%%%%%%%%%%%%%%%%%%%%%%%%%%%%%%%%%%%%%%%%%%%%%%%%%%%%%

\section{Luminosity Functions and Distributions}\label{LumFun}

The luminosity function (LF), defined as the distribution of galaxy luminosities (or magnitudes) per volume, has been for long well studied as a basic statistic. Since galaxy surveys are generally apparent magnitude limited at the faint end, the LF differs from the luminosity distribution (LD) in that the former cannot be obtained from a simple raw histogram of the luminosity data points as LDs are. In fact, the faint luminosity tail of LDs is incomplete, as faint galaxies can be observed only at close enough distances (Malmquist bias). On the contrary, the brightest galaxies can generally be observed over the whole redshift limits of the survey. As a consequence, LFs are identical to LDs at the bright end (except for a scale factor equal to the survey's volume $V_{\rm S}$) but start to depart from each other at a departure luminosity $L_{\rm D}$ (specified next).

The important link between LFs and LDs is that, since they behave the same way at the bright end, we can study LFs in this regime by instead doing the sampling and EVS on the LD of the individual data points. This is the strategy followed in this paper, which works as long as we sample galaxies with luminosities close enough to or brighter than $L_{\rm D}$.

\begin{figure}[h]
\epsscale{1.0}
\plotone{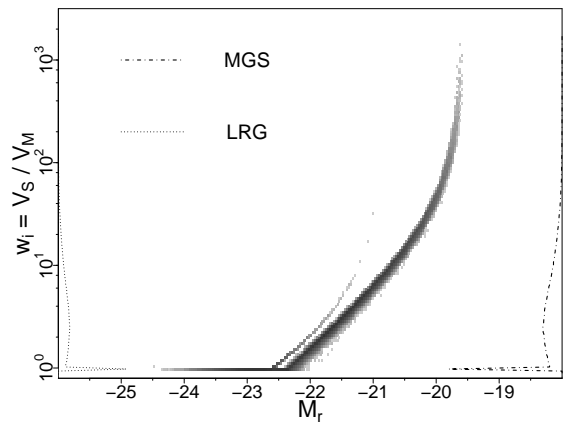}
\caption{2-Dimensional histogram (in logarithmic gray-scale) showing the onset of absolute magnitude incompleteness. For the LRGs (left branch), we see $w_{i} \geq 1$ and $w_{i} \geq 2$ starting at $M_{\rm r}=-22.63$ and $M_{\rm r}= -22.05$). For the MGS (right branch), it happens at $M_{\rm r}=-22.43$ and -21.91, respectively. The curves against the vertical axis are the corresponding (unnormalized) distributions of the weights $w_{i}$.} 
\label{Fig:AbsMagVsVmax}
\end{figure}

In order to correct the incompleteness of low luminosity galaxies, we construct LFs by adding more weight to these galaxies, as used in the Vmax method \citep{schmidt1968}, where each $i$-th galaxy is assigned a weight $w_{i}=V_{\rm S}/V_{{\rm M},i}\geq 1$. Here we note that, given the particular $[z_{1},z_{2}]$ and $[m_{1},m_{2}]$ intervals for the survey, the $i$-th galaxy found at $z_{i}$ could be observed only within a maximum comoving volume $V_{{\rm M},i}$ inside the overall volume $V_{\rm S}$ of the survey. If the $i$-th galaxy of apparent magnitude $m_{i}$,  k-correction $k_{i}=k(z_{i})$, evolution correction $e_{i}=e(z_{i})$ and at a luminosity distance $D_{L}(z_{i})$ were to have limiting apparent magnitudes $m_{1,2}$, then it should be moved to a limiting luminosity distance given by

\begin{eqnarray}
\centering
& D_{L,i}(z_{\rm lim};m_{1,2}) =    \nonumber \\  
& D_{L}(z_{i}) \times  10^{(m_{1,2} - k(z_{\rm lim}) - e(z_{\rm lim}) - m_{i} + k_{i}+ e_{i})/5}. \label{Eq:DlumLim}
\end{eqnarray}
Hence, the maximum volume is defined by the biggest interval of $D_{L}$ inside which a galaxy can appear in the survey:
\begin{eqnarray}
V_{{\rm M},i} & = & [V(\min (D_{L}(z_{2}),D_{L,i}(z_{\rm lim};m_{2})   )) \nonumber \\
					 & - & V(\max (D_{L}(z_{1}),D_{L,i}(z_{\rm lim};m_{1})   ))] \times F_{A},
\label{Eq:Vmax}
\end{eqnarray}

As Eq. \ref{Eq:DlumLim} defines $z_{\rm lim}$ in an implicit way, we solve for it iteratively. 
The weights $V_{\rm S}/V_{{\rm M},i}$ are shown in Fig. \ref{Fig:AbsMagVsVmax}. 
The departure magnitudes separating the complete ($w_{i}=1$) and 
incomplete ($w_{i}>1$) parts of the samples take the values of 
$M_{\rm D}$ = -22.63 (LRG) and $M_{\rm D}$ = -22.43 (MGS).
Thus, the distribution of weights looks bimodal, where the 
complete part of the sample creates the spike at $w_{i}=1$, and 
the incomplete part forms the broad tail. Note that the 
incomplete part presents at the beginning a nearly linear 
trend given by $\log w \sim \frac{3}{5}M_{r}$ (derived from Eq. \ref{Eq:DlumLim}). 
The part of the trend that departs and seems extending into 
the high $\log w$ region, on the other hand, is composed of 
galaxies whose apparent magnitude is very close to the limiting 
apparent magnitude cut $m_{2}$ of the survey.

A non-parametric LF can be then easily estimated using a Vmax weighted histogram in the form

\begin{equation}
\Phi (M) \Delta M = \frac{1}{\Delta M}\sum_{M_{i}\in \Delta M }\frac{c_{i}}{V_{{\rm M},i}}
\label{eq:LumFunVmax},
\end{equation}
The extra weight $c_{i}$ takes into account the incompleteness of the target selection algorithm for spectroscopic follow up. The error $\delta \Phi (M_{r})$ is estimated by using JackKnife sampling of 38 regions about 200 SqDeg each.

\begin{figure}[t]
\epsscale{1.15}
\plotone{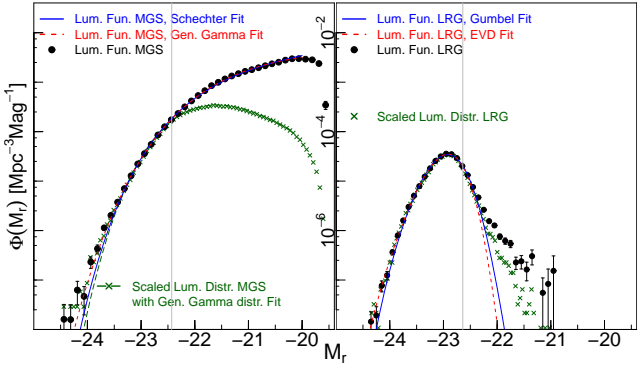}
\caption{Luminosity functions and distributions of the different galaxy samples, together with the best Schechter, generalized gamma, Gumbel and EVD fits from Tables \ref{table:FittingParametersLumFunction} and \ref{table:FittingParametersLumDistribution}. The vertical lines denote the completeness boundary $w_{i}=1$, where the LFs depart from LDs. The LDs were scaled by the factor $N_{g}/V_{\rm{S}}$ in order to compare them with the LFs.}
\label{fig:LumFun}
\end{figure}

The luminosity functions and distributions and their fits are shown in Fig. \ref{fig:LumFun}, and Tables \ref{table:FittingParametersLumFunction} and \ref{table:FittingParametersLumDistribution} contain the fitting parameters. 

For the MGS samples, we use a generalized gamma as a fitting function, which results in the known Schechter profile \citep{schechter1976} when $\beta=1$ : 
\begin{equation}
\Phi(L)dL = \Phi_{*} \left( L/L_{*} \right)^\alpha \exp \lbrace -\left[ (L/L_{*})^{\beta} \right] \rbrace dL.
\label{eq:SchechterFunction}
\end{equation}
The normalization factor $\Phi_{*}$ is left as a free parameter for LFs, whereas for the LDs is defined by $\Phi_{*}=\beta/[L_{*}\Gamma((\alpha+1)/\beta,(L_{{\rm min}}/L_{*})^{\beta})]$, where the incomplete gamma function $\Gamma( \ast, (L_{\rm min}/L_{*})^{\beta} )$ is the integral in the interval $[(L_{\rm min}/L_{*})^{\beta},\infty]$.     
The fitted luminosity values for the LFs and LDs we converted into magnitude units using $M_{r}$=$-2.5\log_{\rm 10} [L/L_{\odot}]$+$M_{\odot,r}$, where $M_{\odot,r}=4.62$ \citep{blanton2001} and $L_{\odot}$ is the solar luminosity in the $r$ band. 

From Figure \ref{fig:LumFun}, we can see that the generalized gamma function provides a better fitting for the MGS LF than the Schechter fit. Indeed, $\beta=0.75$ fits much better the high luminosity tail, and is similar to the value found by \cite{bernardi2010} ($\beta=0.698$). Our faint end slope ($\alpha=-0.81$) is steeper compared to their value ($\alpha=-0.45$), although we are fitting in a different magnitude interval and to a galaxy sample of different magnitude and redshift selection cuts. The errors in the magnitude have little influence in the value of the fitted parameters. As shown in \cite{bernardi2010}, bigger magnitude errors might decrease the fitted value of $\beta$. However, they showed that the inclusion of the $\lesssim$0.05 rms errors on the magnitude in SDSS provided discrepancies in the fitting parameters generally smaller than their statistical errors.

The LRG sample was made to include the brightest early types. Therefore, they are naturally better fitted with an extreme value distribution (EVD) or its special case the Gumbel where $\xi=0$ (see Sec. \ref{section:TheoryOfEVS}):

\begin{eqnarray}
& \Phi(L)=  \Phi_{*} {\rm EVD}(L), &   \label{Eq:EVD} \\
& {\rm EVD}(L)=  L_{\sigma}^{-1} t(L)^{\xi+1} \exp(-t(L))dL, &    \label{eq:GEVluminosity} 																						 \\
& t(L) =    ( 1+ \xi [ \frac{L-L_{\mu}}{L_{\sigma}}     ] )^{-1/\xi},   1+ \xi [ \frac{L-L_{\mu}}{L_{\sigma}}]  > 0. &  \nonumber 
\end{eqnarray}
with $L_{\mu}$, $L_{\sigma}$ and $\xi$ being respectively the location, scale and shape (or tail index) fitting parameters. The LRGs were built to be a complete (volume limited) sample, but some scattered lower luminosity galaxies passed the color cuts and contaminated it (Fig. \ref{Fig:AbsMagVSredshift}). Therefore, we only fit the LF up to the completeness limit $M_{\rm D} = -22.64$ as explained earlier. 

\begin{table}{}
\begin{center}
\caption{Luminosity Function Fitting parameters.\rlap{$^a$}}
\scriptsize \addtolength{\tabcolsep}{-2pt}
\begin{tabular}{ccccc}
\hline
\hline
Sample 	&	$\Phi_{*}{\tiny [10^{-3}}$							&	$M_{*}$	&	$\alpha$	& $\beta$  \\
			&	${\tiny {\rm Mpc^{-3}Mag^{-1}}]}$	&				&				\\
\hline
\hline
 &		{\small Schechter} & {\small Fitting} & &  \\ 
\hline
MGS 	&			3.10 $\pm$ 0.05	&			-21.46 $\pm$ 0.02		&			-1.34 $\pm$ 0.04		& 1 \\

\hline
	&	{\small General} & {\small Gamma}  & {\small Fitting} & \\ 
\hline
MGS 	&			7.79 $\pm$ 0.38	&			-20.42 $\pm$ 0.10		&			-0.81 $\pm$ 0.05		&  0.75 $\pm$ 0.02  \\
\hline
\hline
Sample 	&	$\Phi_{*}$							&	$M_{\mu}$	&	$M_{\sigma}$	&	$\xi$ \\ 
			&	$[10^{-5}\,{\rm Mpc^{-3}}]$	&				&				&        \\
\hline
\hline
	&	{\small Gumbel} & {\small Fitting}  &  & \\ 
\hline
LRG 	&				2.52$\pm$0.03		&			-22.85$\pm$0.01			&	-21.36$\pm$0.02			& 0	\\
\hline
	&	{\small GEV} & {\small Fitting}  &  & \\ 
\hline
LRG  	&				2.49$\pm$0.02		&			-22.86$\pm$0.01			&	-21.36$\pm$0.02			&	0.04$\pm$0.01\\

\hline
\end{tabular}
\label{table:FittingParametersLumFunction}
\end{center}
{\small $^a$~ Parameters from fitting to Eq. \ref{eq:SchechterFunction} (MGS) or Eq. \ref{Eq:EVD} (LRG). Parameters in luminosity units ($L_{*}$, $L_{\mu}$ and $L_{\sigma}$) were converted into absolute magnitudes  ($M_{*}$, $M_{\mu}$ and $M_{\sigma}$) using  $M$=$-2.5\log_{\rm 10} [L/L_{\odot}]$+$M_{\odot}$, where $M_{\odot}=4.62$, everything measured in the petrosian r-band. MGS and LRG samples are fitted in the ranges $M_{r}\leq -20.2$ and $M_{r}\leq -22.64$ respectively.}
\end{table}

\begin{table}{}
\begin{center}
\caption{Luminosity Distribution Fitting parameters.\rlap{$^a$}}
\scriptsize \addtolength{\tabcolsep}{-2pt}
\begin{tabular}{cccc}
\hline
\hline
Sample 	&		$M_{*}$	&	$\alpha$	& $\beta$  \\
			&					&				& \\
\hline
\hline
	&	{\small General} & {\small Gamma}  & {\small Fitting}  \\ 
\hline
MGS 	&						-19.99 $\pm$ 0.16		&			1.52 $\pm$ 0.10		&  0.79 $\pm$ 0.03  \\
\hline

\hline
\end{tabular}
\label{table:FittingParametersLumDistribution}
\end{center}
{\small $^a$~ Parameters from fitting to Eq. \ref{eq:SchechterFunction} (MGS) using $\Phi_{*}=\beta/(L_{*}\Gamma((\alpha+1)/\beta,(L_{\rm min}/L_{*})^{\beta}))$. We used $M_{\rm max}\equiv M(L_{\rm min})=-20.2$. Parameters in luminosity units were converted into absolute magnitudes.}
\end{table}

%%%%%%%%%%%%%%%%%%%%%%%%%%%%%%%%%%%%%%%%%%%%%%%%%%%%%%%%%%%%%%%%%%%%%%%%%%%%%%%%%%%%%%%%%%%%%%%%%%%%%%%%%%%%%%%%%%%%%%%%%%%%%%%%%%%%%%%%%%%%%%%%%%%%%%%%%%%%%%%%%%%%%%%%%%%%%%%%%%%%
%%%%%%%%%%%%%%%%%%%%%%%%%%%%%%%%%%%%%%%%%%%%%%%%%%%%%%%%%%%%%%%%%%%%%%%%%%%%%%%%%%%%%%%%%%%%%%%%%%%%%%%%%%%%%%%%%%%%%%%%%%%%%%%%%%%%%%%%%%%%%%%%%%%%%%%%%%%%%%%%%%%%%%%%%%%%%%%%%%%%

\section{Sampling the Maximal Luminosities: Creation of i.i.d. batches and HEALPix-based pencil beams}\label{section:FootPrintCreation}

Classic extreme value statistics needs close-to-i.i.d realizations of the underlying parent probability distribution from which to draw the maximal values.

In the EVS of time series, a common practice is to use the block maxima approach, where the (possibly correlated) data set is grouped into 
disjoint and temporally consecutive blocks or batches of the same size from which to choose the extreme values (e.g. annual maxima) \citep{embrechts1997, reiss1997, coles2001}. 
The blocks can be chosen to span each of the many different cycles of the underlying process, 
generally creating $n$ blocks, with all the blocks having the same number $N$ (batch size) of data points.
Thus, a first simplified sampling strategy in this paper is the one where the disjoint batches are chosen by random sampling without replacement of the luminosity values.

In real life situations, however, some blocks might present missing or sparse data, due to bad sampling strategies or sensor failures. 
In other cases, the data points clump in clusters of different sizes that exceed a certain threshold level (e.g., insurance claims after a hurricane). 
In all these situations, the different realizations of the parent probability distribution will have different number $N$ of (possibly correlated) data points, with distribution P($N$). 

\begin{figure}[ht]
\begin{center}
  \includegraphics[width=3.5cm,angle=90]{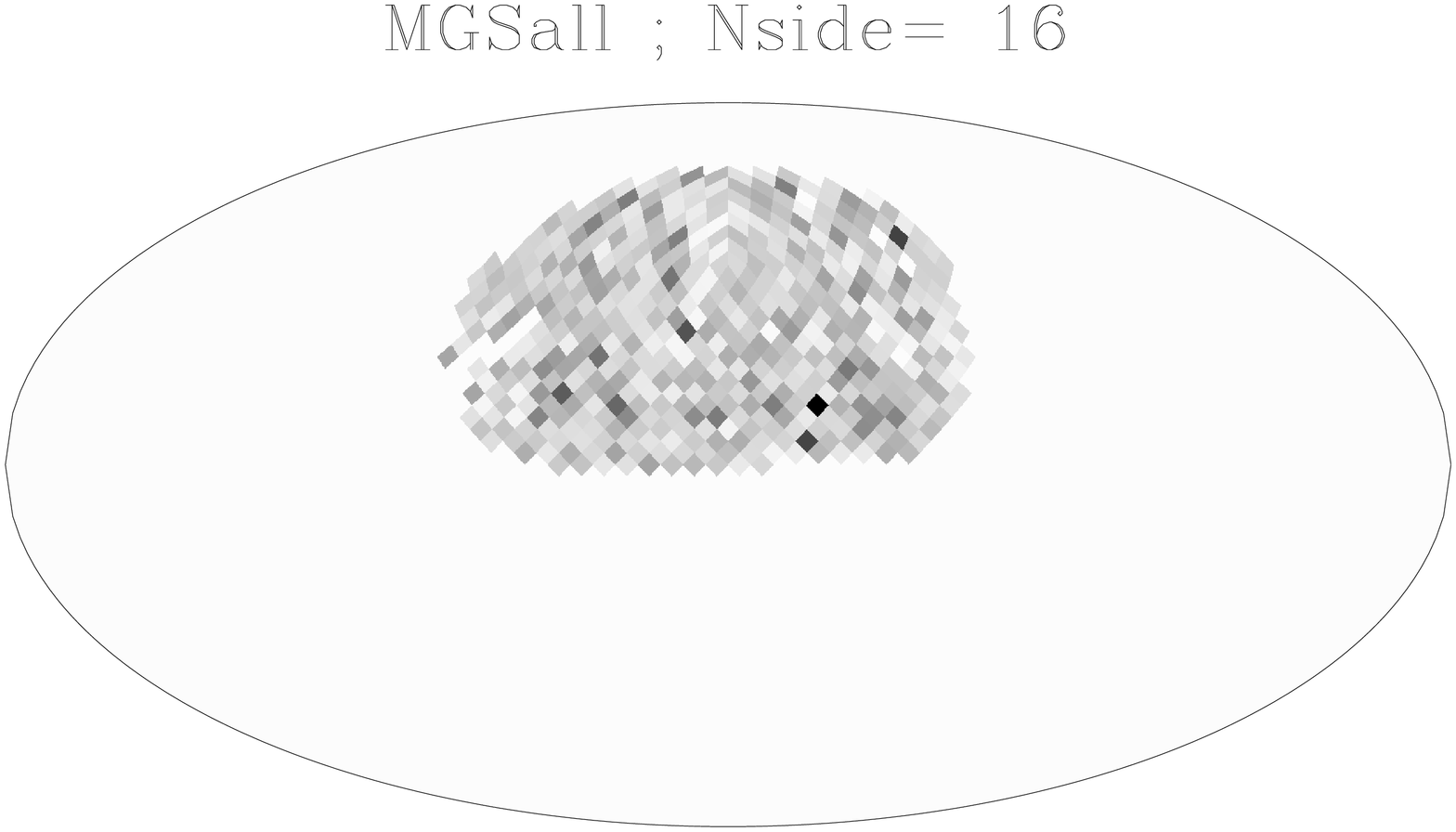}\\
  \includegraphics[width=3.5cm,angle=90]{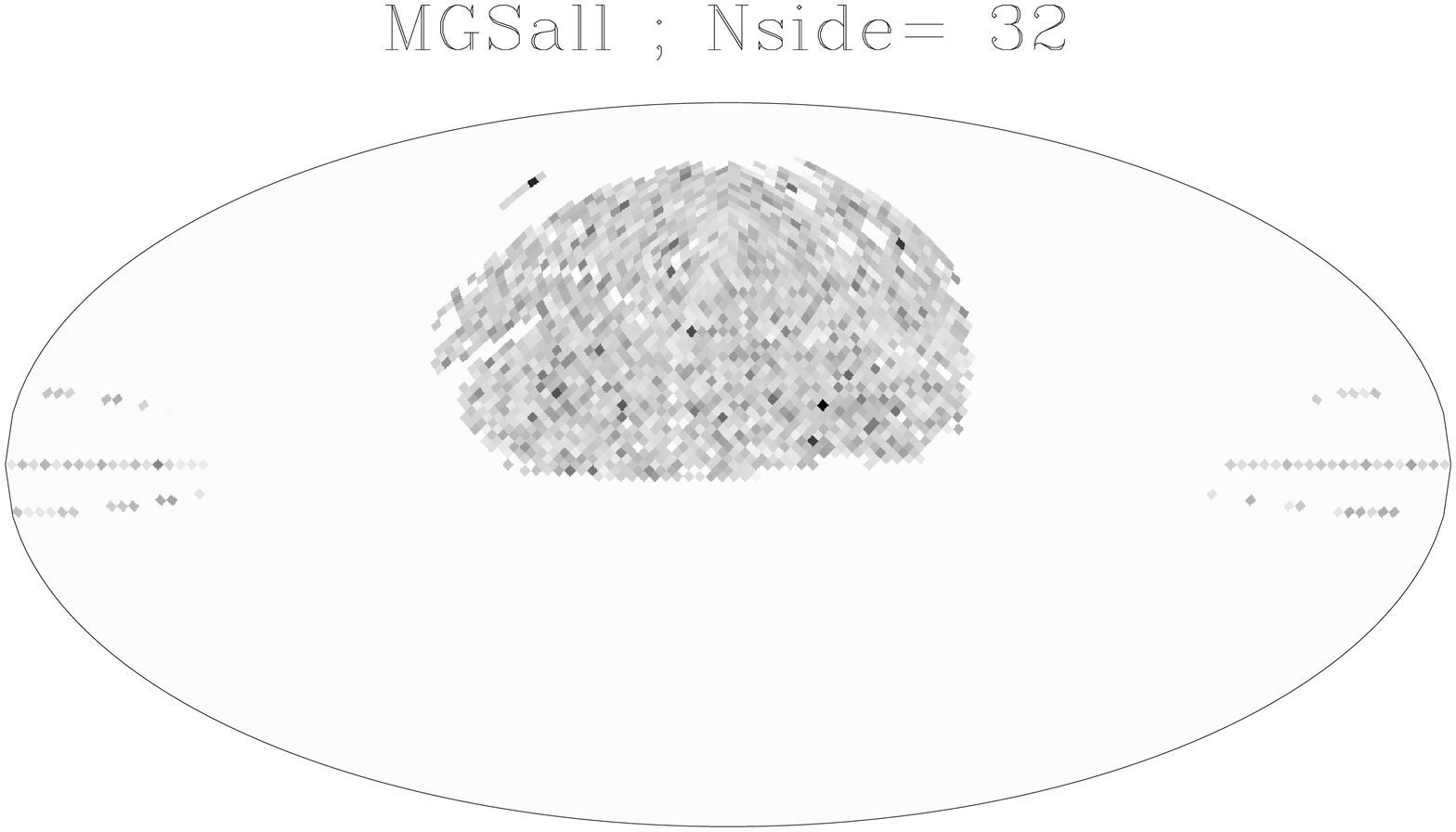}\\
  \includegraphics[width=3.5cm,angle=90]{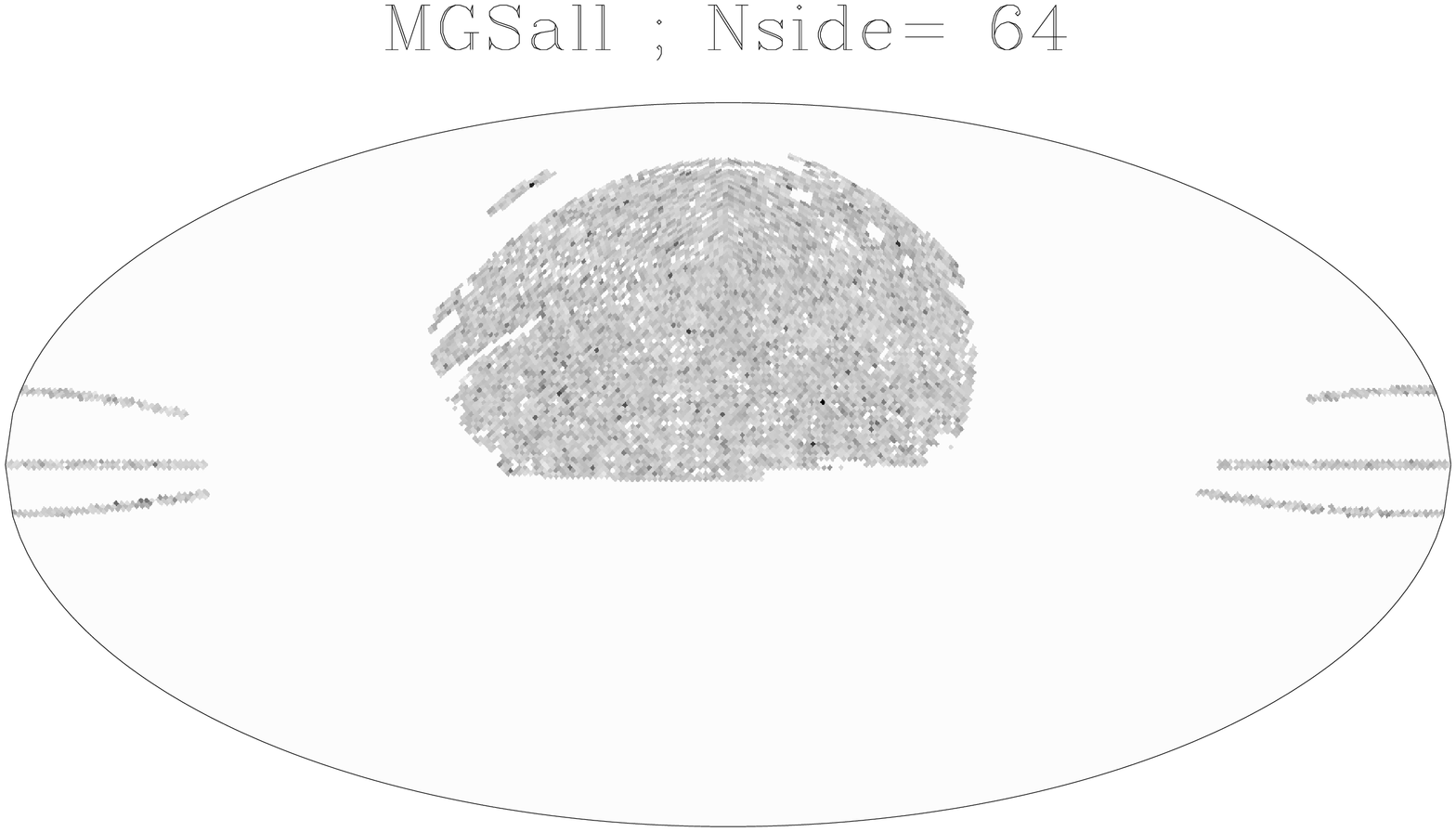}\\
\end{center}
\caption{HEALPix Maps of maximal luminosities (in linear scale) for the MGS galaxy sample at different values of $N_{side}$. Darker color means higher luminosity. Only cells included 97\% in the footprint are shown. The SDSS-DR8 footprint boundaries become evident at resolution $N_{side}=64$.}
\label{fig:HEALPixMaps}
\end{figure}

\begin{figure}[h]
\epsscale{1.0}
\plotone{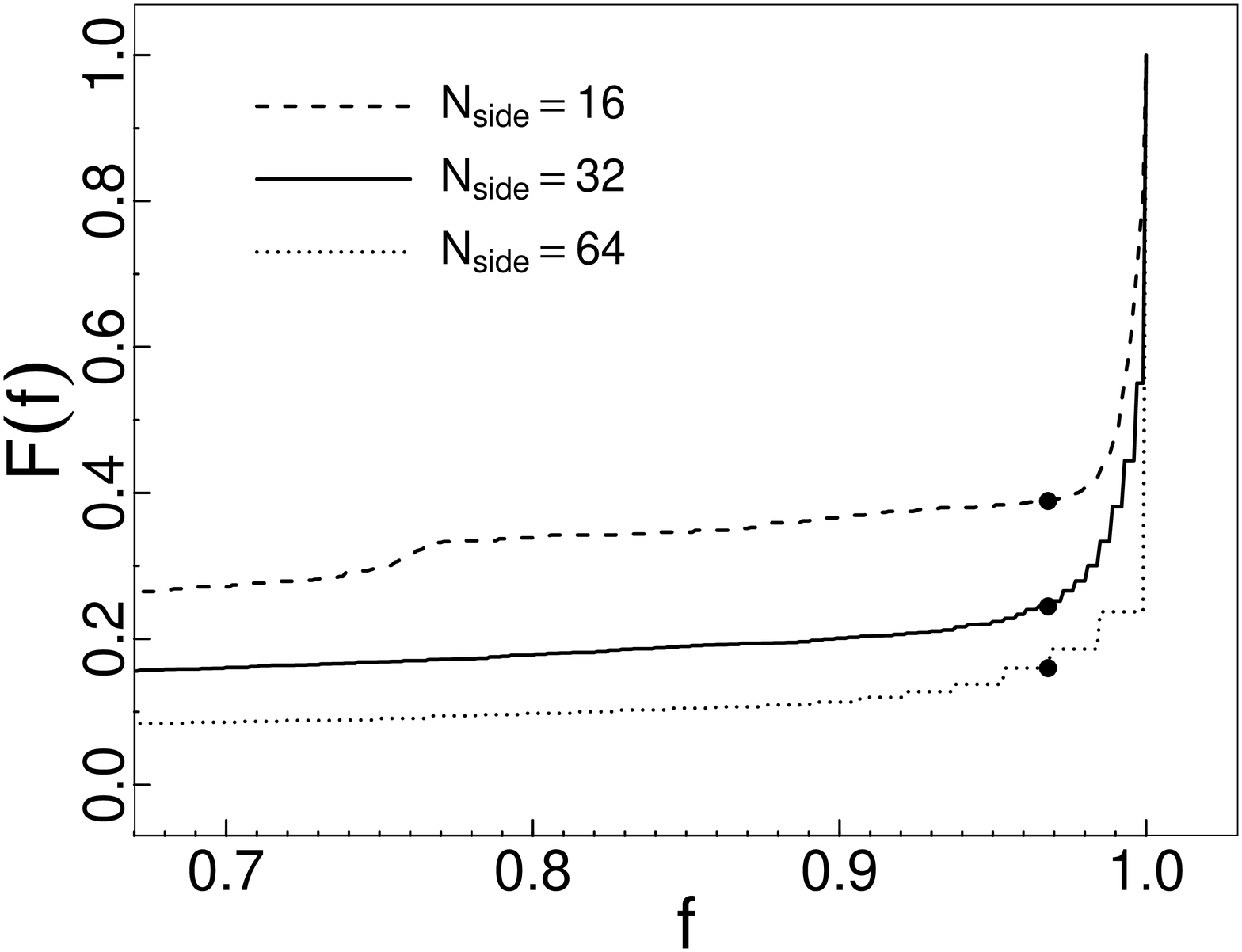}
\caption{Cumulative distribution of the HEALPix cell fractional occupation area $f$, shown for the 3 different footprint cell sizes. The black filled dots show the breakpoints at $f=0.97$. We consider only cells with $f\geq 0.97$}
\label{fig:CumulativeOfF}
\end{figure}

In order to show a real working equivalent example of the previous time series process, the second sampling strategy generalizes the block maxima approach by extending it to the case of variable block size and weak enough correlations (the meaning of $weak$ is discussed in Section \ref{disc}). To this aim, we recreate this situation by dividing the sky in equal-area patches, each one defined by an individual cell of the HEALPix tessellation \citep{gorski2005}. This creates 1-dimensional pencil-like beams, each of which containing one close-to-i.i.d realization of the galaxy distribution through redshift and with a variable galaxy number $N$. 
We start with a finer SDSS DR8 spectroscopic footprint with resolution $N_{side}=512$ (of cell size $\sqrt{\Omega_{pix}} \simeq 6.87'$).
We further degrade the footprint into 3 lower resolution maps defined by $N_{side}=16,32$ and $64$, creating thus the cells that define the pencil beams. Note that these bigger cells may partly cover an area not belonging to the footprint. Hence, we define the fractional area occupancy $f$ as the area inside the footprint covered by the cell divided by the total area of the cell. The cumulative distribution of $f$ (Fig. \ref{fig:CumulativeOfF}) shows clear breakpoints at $f \sim 0.97$ for all 3 resolutions. We therefore decide to use only the group of cells which satisfy $f\geq 0.97$. A summary on the 3 different resolution HEALPix schemas is presented in Table \ref{table:HEALPixAndNumPix}, and HEALPix maximal luminosity maps are shown for the MGS in Fig. \ref{fig:HEALPixMaps}.
\newline

\begin{table} 
\caption{HEALPix schemas for the different galaxy samples \rlap{$^a$}}
\label{table:HEALPixAndNumPix}
\begin{center}
%\begin{tabular}{||c||c||c||c|c|c|c||c|c|c|c||}\hline\hline 
\begin{tabular}{ccc|cccc}\hline\hline 
\multicolumn{7}{c}{HEALPix schemas} \\ 
\hline\hline
%$N_{side}$ & $\sqrt{\Omega_{pix}}$	&	$n_{sphere}$	&		\multicolumn{4}{|c||}{$n_{F}$}		& \multicolumn{4}{|c||}{$n \equiv n_{F,97}$}   \\
$N_{side}$ & $\sqrt{\Omega_{pix}}$	&	$n_{sphere}$	&		$n_{F}$	&  $n_{F}$ & $n_{F,97}$ & $n_{F,97}$  \\
\cline{4-7} 
			  & 								& 						&	MGS	&	LRG	& MGS	&	LRG \\ 
\hline
16 & $3.66^{o}$ &  3072 &  768 	&  755 &  473 &  473 \\
32 & $1.83^{o}$ & 12288 & 2659 	& 2591 & 2030 & 2029 \\
64 & $55.0'   $ & 49152 & 10017	& 9461 & 8492 & 8256 \\
\hline\hline
\end{tabular}
\end{center}
{\small
$^a$~  Here, $n_{sphere}$ is the total number of HEALPix cells in the sky (each of area $\Omega_{pix}$), $n_{F}$ is the number of galaxy-containing cells inside the footprint and $n_{F,97}$ is the number of cells with areas included at least $97\%$ inside the footprint.}
\end{table}

\subsection{Distributions of galaxy counts in a HEALPix cell}\label{Sec:BatchSizeDistribution}

\begin{figure*}[htb]
\begin{center}
\includegraphics*[width=5.6cm, height=4.6cm]{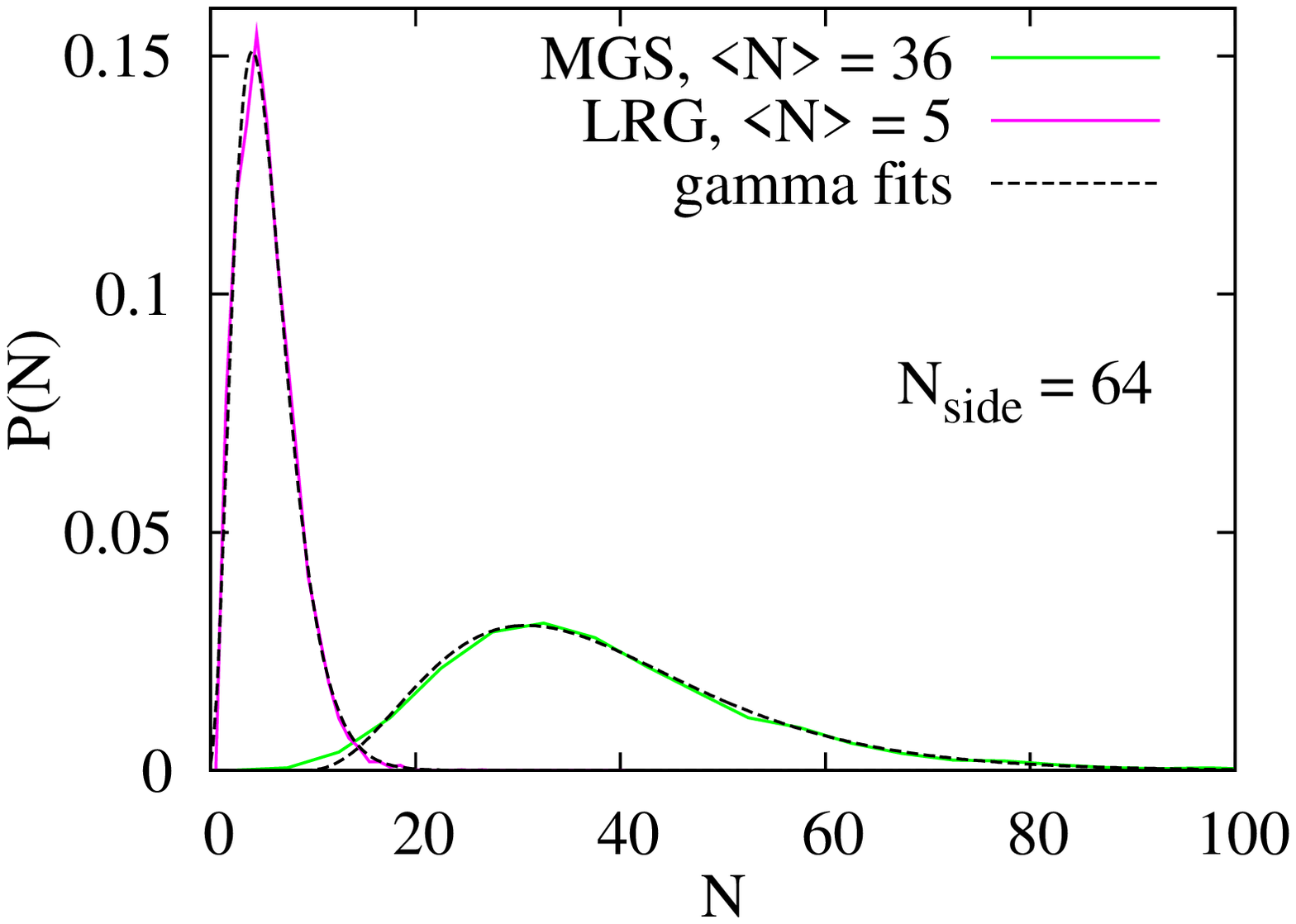}
\includegraphics*[width=5.6cm, height=4.6cm]{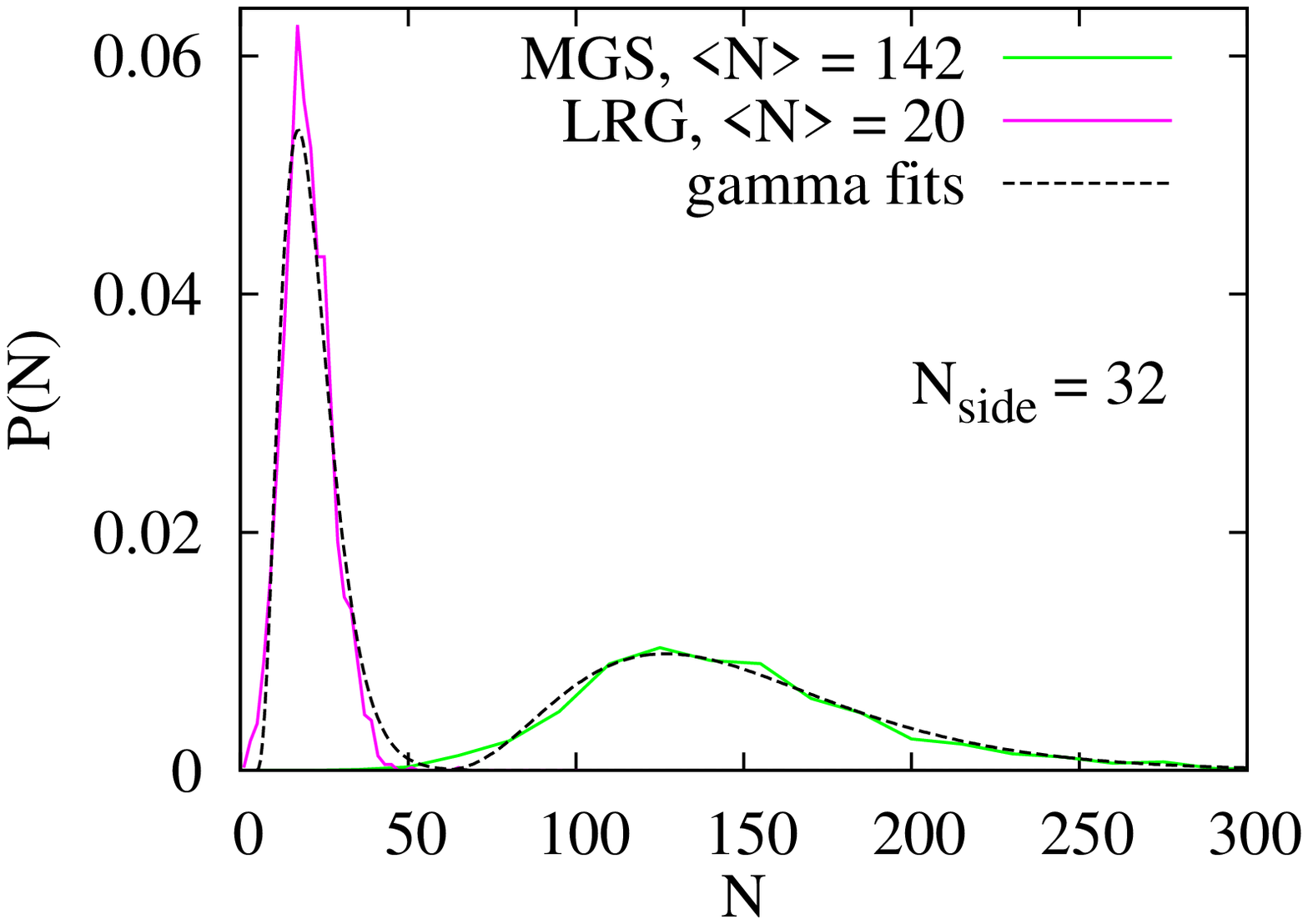}
\includegraphics*[width=5.6cm, height=4.6cm]{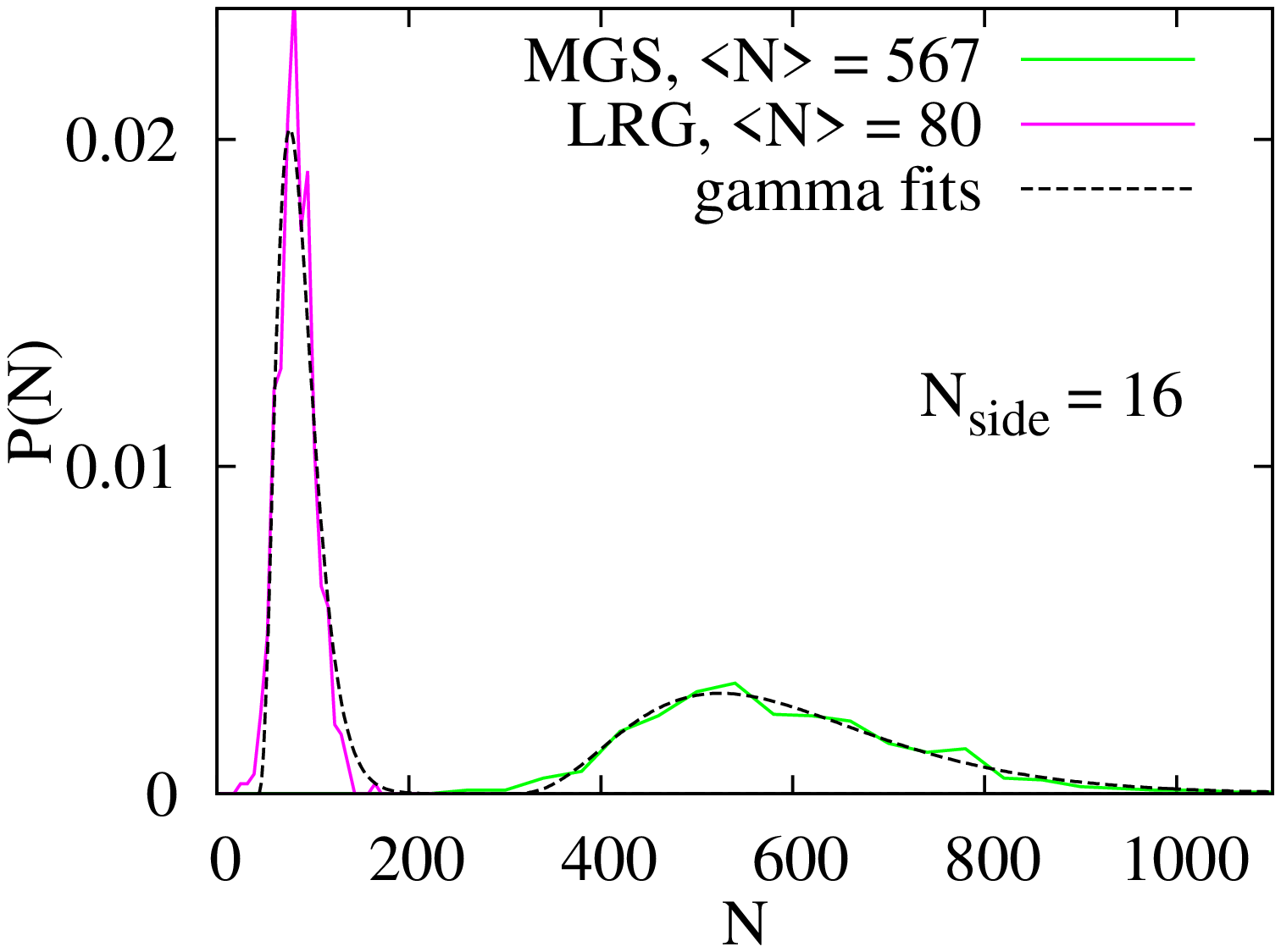}\\
\includegraphics*[width=5.6cm, height=4.6cm]{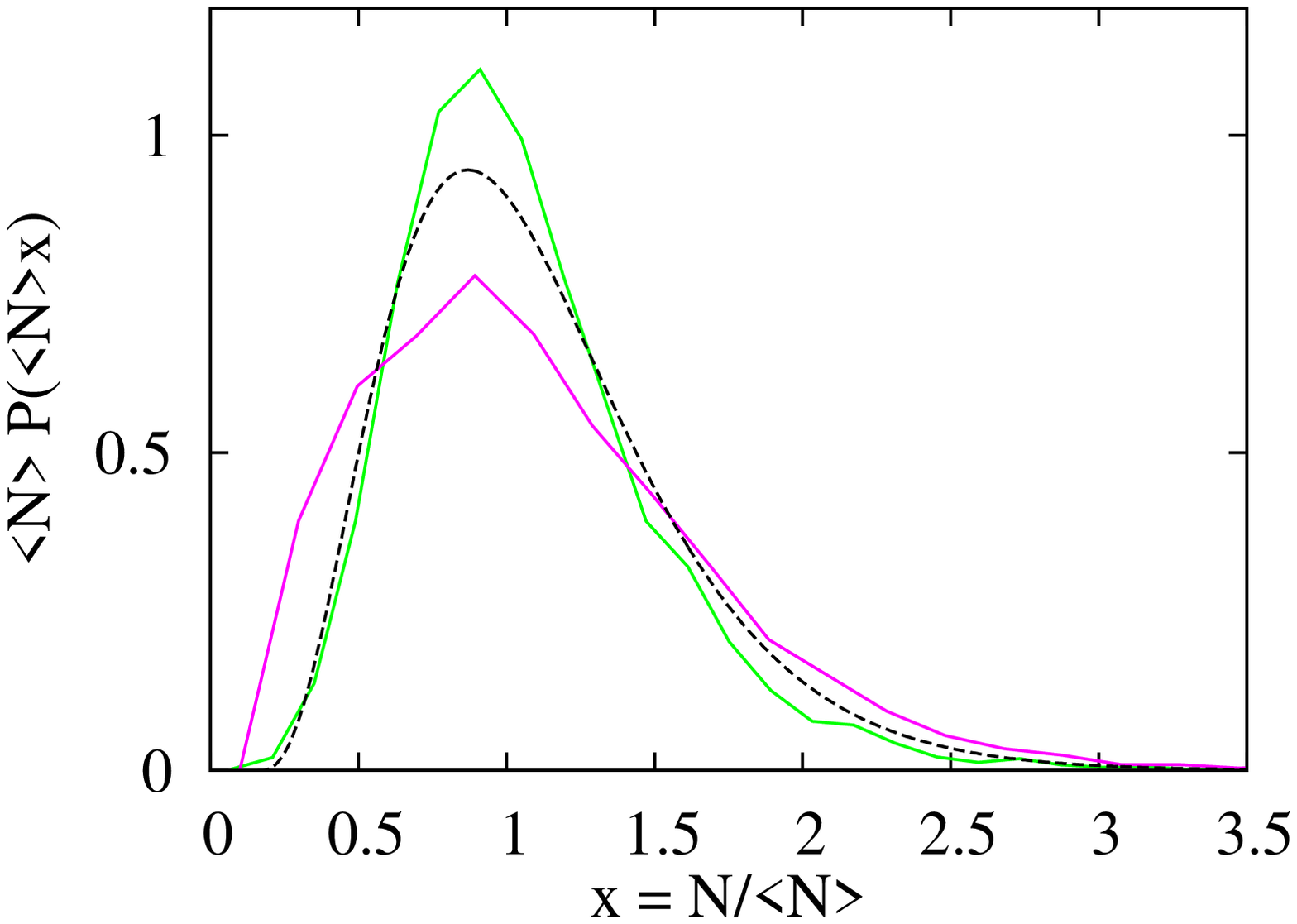}
\includegraphics*[width=5.6cm, height=4.6cm]{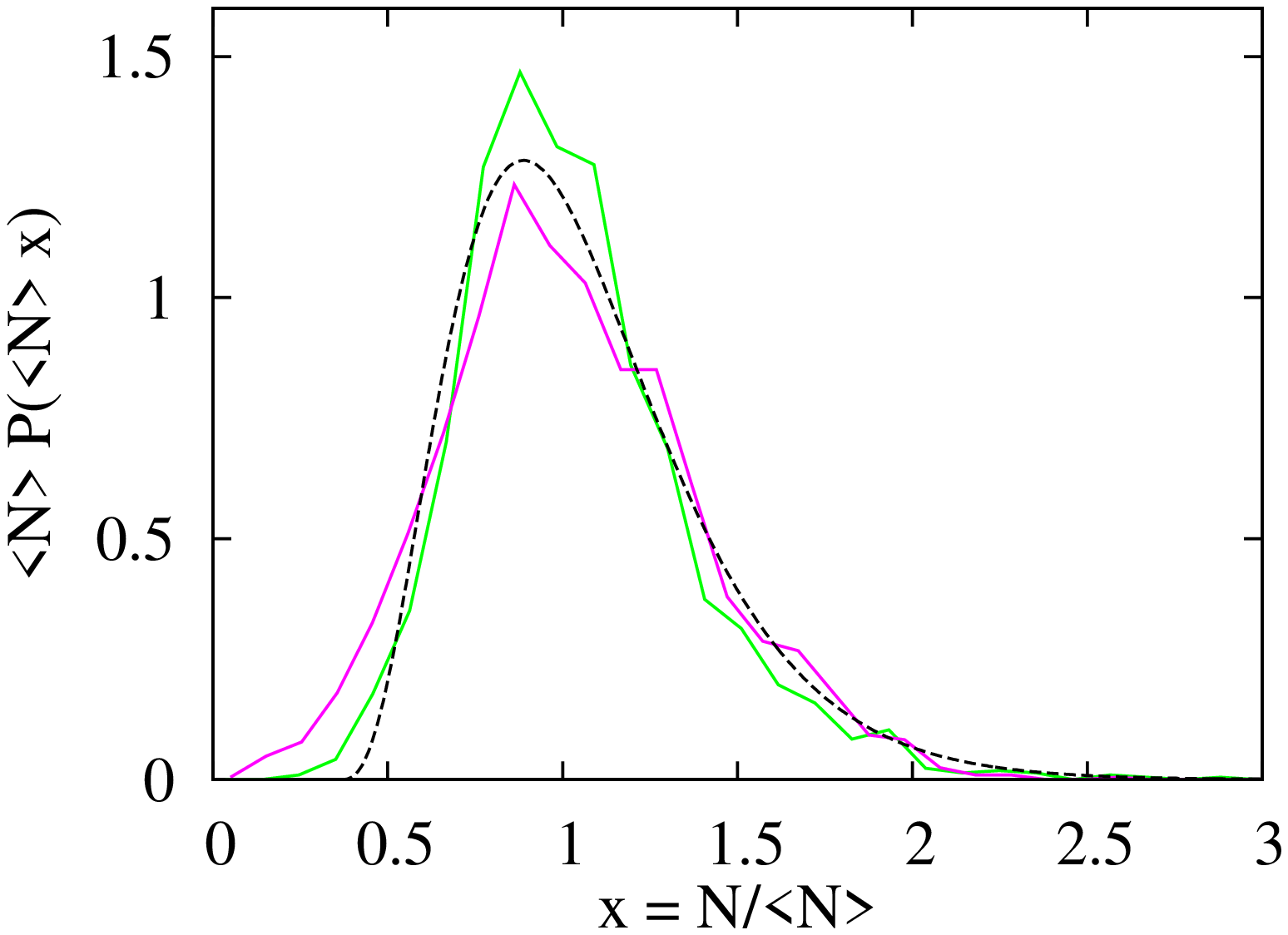}
\includegraphics*[width=5.6cm, height=4.6cm]{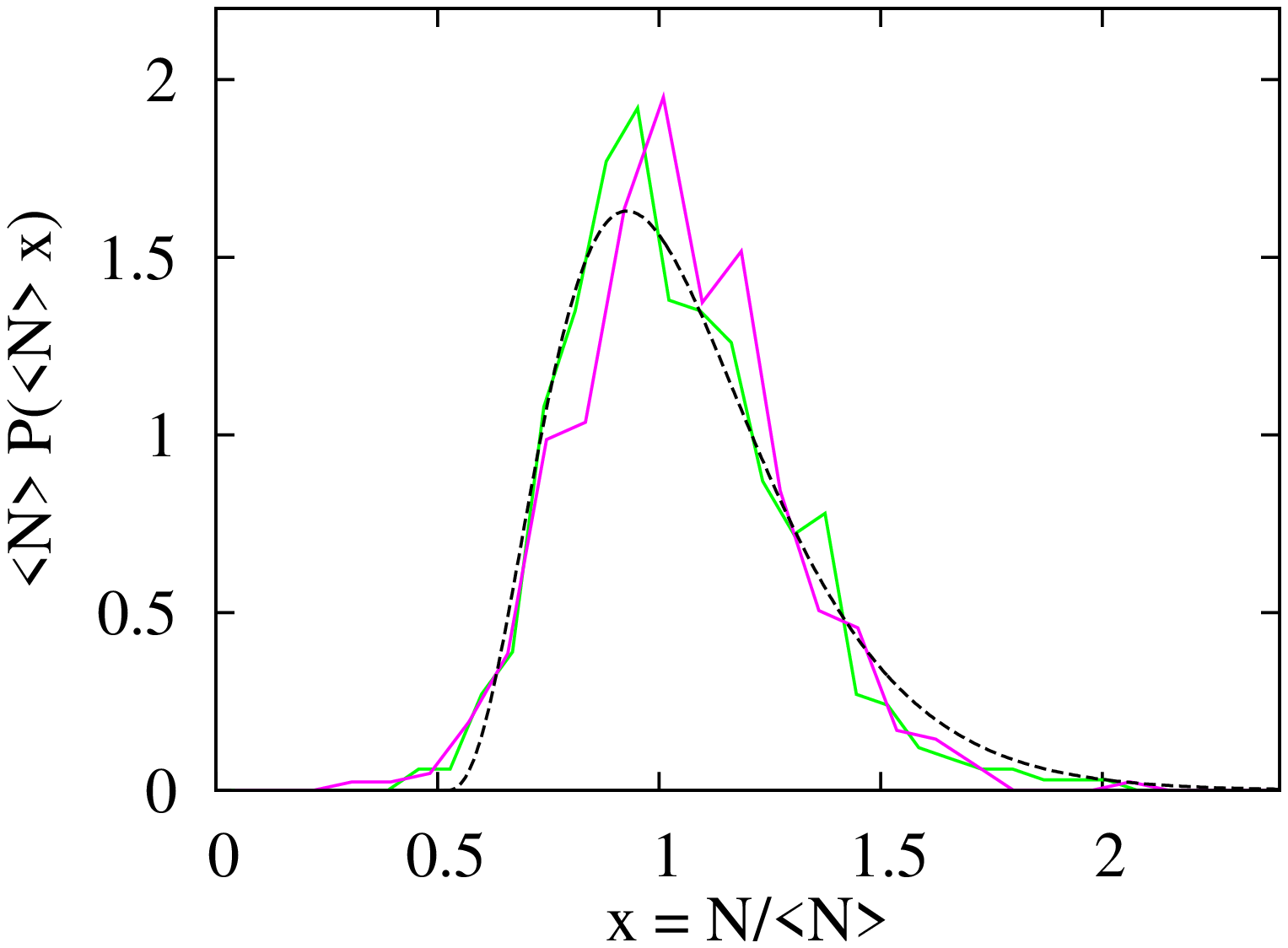}
\end{center}
\caption{Galaxy count distributions in a HEALPix cell of size $N_{side}$ = 64, 32, and 16, i.e. for increasing average galaxy count $\langle N\rangle$. Gamma function of the form ${\cal F}(N)$ $\sim$ $(N/\overline{N}-d)^3$ $\exp(-N/\overline{N} + d)$ fits well the distributions as can be seen and, to a good approximation, count distributions for different galaxy samples can be scaled together by the average number of galaxies in a cell $\langle N\rangle$ (second row). Note that equally good fits can also be provided by negative binomial distributions used in previous studies, e.g. \cite{croton2007}, \cite{yang2011} and references therein.}
\label{Number-distr}
\end{figure*}

The number of galaxies $N$ fluctuates in the pencil beams. Unless the galaxy-count distribution in a HEALPix cell ${\cal F}(N)$ is very narrow, this affects the limit distribution one expects for the extreme luminosities. We have thus evaluated ${\cal F}(N)$ for the MGS and LRG samples, and the results for decreasing $N_{side}$, i.e. for increasing average of the galaxy count $\langle N\rangle$ are shown in Fig.\ref{Number-distr}.

As one can see, the $N_{side}=64$ suggest good statistics yielding smooth functions. We should note, however, that the average galaxy count is rather low in this case ($\langle N\rangle\approx 5-40$), so it is far from the limit $\langle N\rangle\to\infty$ one would like to take when investigating the EVS.

There are two lessons to learn from Fig.\ref{Number-distr}. First, the distributions are rather narrow which suggests that it is a reasonable assumption that the theory of EVS known for fixed N can be applied to galaxy luminosity. Second, one can develop analytic approximations to these histograms. Indeed, the distributions can be relatively well approximated by a gamma function with free location and scale parameters. 

%%%%%%%%%%%%%%%%%%%%%%%%%%%%%%%%%%%%%%%%%%%%%%%%%%%%%%%%%%%%%%%%%%%%%%%%%%%%%%%%%%%%%%%%%%%%%%%%%%%%%%%%%%%%%%%%%%%%%%%%%%%%%%%%%%%%%%%%%%%%%%%%%%%%%%%%%%%%%%%%%%%%%%%%%%%%%%%%%%%%
%%%%%%%%%%%%%%%%%%%%%%%%%%%%%%%%%%%%%%%%%%%%%%%%%%%%%%%%%%%%%%%%%%%%%%%%%%%%%%%%%%%%%%%%%%%%%%%%%%%%%%%%%%%%%%%%%%%%%%%%%%%%%%%%%%%%%%%%%%%%%%%%%%%%%%%%%%%%%%%%%%%%%%%%%%%%%%%%%%%%

\section{Theory of Extreme Value Statistics}\label{section:TheoryOfEVS} 
\subsection{Classical Theory}\label{Sec:ClassicalTheory} 

Extreme value statistics (EVS) is concerned with the probability, $P_N(v)dv$, of the largest value in a batch of $N$ measurements $\{v_1,v_2,...,v_N\}$ being $v=\max_i v_i$. 
For us, the $v_i$s are galaxy luminosities, obtained by either random sampling $N$ galaxies from the sky, or chosen from the HEALPix cells covering the sky, each with a variable $N$.

The results of the EVS are simple provided the $v_i$s are i.i.d. variables drawn from a general parent distribution $f(v_i)$. Namely, the limit distribution $P_{N\to\infty}(v)$ belongs to one of three types and the determining factor is the large-argument tail of the parent distribution \citep{gumbel,galambos}. Frechet type distribution emerges if $f$ decays as a power law, Fisher-Tippett-Gumbel (FTG) distribution is generated by $f$s which decays faster than any power law and, finally, parent distributions with finite cutoff and power law behavior around the cutoff yield the Weibull distribution \citep{gumbel,galambos}. All the above cases can be unified as a generalized EVD whose integrated distribution $F_N(v)$ is given in the $N\to\infty$ limit by
\begin{equation}
F(v)=\exp{ \left\{  - \left[ 1+\xi \left( \frac{v-\mu}{s} \right) \right]  ^{-1/\xi} \right\}   }\label{Eq:CumulativeEVD}  
\end{equation}
with parameters $\mu,s$ and $\xi$. The shape parameter $\xi$ can take values $\xi>0,\, =0,\, <0$, which correspond to the Frechet, FTG, Weibull classes, respectively. The parameter $\xi$ is also called the tail index, since it is related to the exponent of the large-argument power-law behavior. The probability density function associated to Eq. \eqref{Eq:CumulativeEVD} is shown in Eq. \eqref{eq:GEVluminosity}.

The EVS has been developed mainly for i.i.d. variables and there are only a few well established results for systems with correlations between the $v_i$s. These results are mainly related to sufficiently weakly correlated variables where the i.i.d. results can be shown to apply \citep{Berman,Gyorgyi3}. In the following we shall assume that the correlations between the galaxy luminosities are sufficiently weak so that the experimental histograms can be compared with the i.i.d. results. This assumption is important for the sampling in HEALPix cells in the sky, but not in a random sampling schema (arguments in favor of this assumption will be discussed in Section \ref{disc} using the knowledge of the correlations between galaxy positions).

The parent distribution for galaxy luminosities is known to be well fitted by the Gamma-Schechter distribution $\Phi(L) = \Phi_*(L/L_*)^{\alpha}\exp[-(L/L_*)^{\beta}]$ as given in \eqref{eq:SchechterFunction}
where $L^*$ sets the scale, and $\alpha\approx -1$ together with $\beta = 1$ is the Schechter profile. For this parent distribution, the theory of EVS tells us that the limit distribution of extremal luminosities belongs to the FTG class ($\xi\to 0$)
\begin{equation}
P(v)=\frac{dF(v)}{dv}= a\exp\left[-(av+b)-e^{-(av+b)}\right]\, 
\label{FTGlimit}.
\end{equation}
where the parameters can be fixed by setting $\langle v\rangle =0$ and $\sigma =\sqrt{\langle v^2\rangle -\langle v\rangle^2}=1$, yielding $a=\pi /\sqrt{6}$ and $b=\gamma_E\approx 0.577$. It should be emphasized that this choice leads to a parameter-free comparison with the empirical data. In fact, the histogram of the maximal luminosities $P(v)$ should be plotted in terms of the variable $x=(v-\langle v\rangle_N)/\sigma_N$ where $\langle v \rangle_N$ is the average of the maximal luminosity while $\sigma_N =\sqrt{\langle v^2\rangle_N -\langle v\rangle_N^2}$ is its standard deviation. The resulting scaling function should approach the universal function (\ref{FTGlimit}) in the $N\to\infty$ limit 
\begin{equation}
P_N(x)=\sigma_N P_N(\sigma_N x +\langle v\rangle_N)
\to P(x)\, .
\end{equation}

\subsection{Deviations from the Classical Theory}

In addition to the assumption of $v_i$s being i.i.d. variables, there are two additional problems with the program of comparing the data with the theory. First, a notorious aspect of EVS is the slow convergence of $P_N(x)$ to the limit distribution $P(x)$. Second, the batch size N (the number of galaxies in a given solid angle) varies with the direction of the angle. Thus the histogram of the maximal luminosities $P_N(x)$ is built from a distribution of $N$s. Both of the above effects introduce corrections to the limit distribution we are trying to use for comparison. Below we estimate the magnitude 
of these corrections.
%\newline

\subsubsection{Finite size corrections}\label{Sec:FiniteSizeCorrection}

Finite size corrections in EVS have been studied in detail with the main conclusion that to first order in the vanishing correction in the $N\to \infty$ limit, the scaling function can be written as
\begin{equation}
P_N(x)\approx P(x) + q(N)P_1(x) \, , \label{Eq:P_N(x)}
\end{equation}
where $q(N\to\infty)\to 0$ and the shape correction $P_1(x)$ is universal function. Both the amplitude $q$ and the shape correction $P_1(x)$ are known for Schechter type parent distributions. The convergence to the limit distribution is slow since we have \citep{Gyorgyi4,Gyorgyi5}
 \begin{equation}
q(N)=-\frac{\alpha}{\ln^2{N}} \, .
\end{equation}
for $\beta=1$. In the general case of a parent following the generalized gamma distribution of Eq. \eqref{eq:SchechterFunction}, with $\beta\approx 1$, there are two terms which may have comparable contributions (with the shape correction function being identical)
 \begin{equation}
q(N)= \frac{(1-\beta)}{\beta \ln{N}} +  \frac{(2\beta -1)(\beta-\alpha-1)}{\beta^2 \ln^2{N}} . \label{qNforGenGamma}
\end{equation}
Note that this theoretical construct needs the values of $\alpha$ and $\beta$ to be fitted at the bright end tail of the luminosity distribution, thus neglecting the low luminosity tail \citep{Gyorgyi5}. The value of $\beta$ is roughly $1$, thus for a characteristic range of $N\approx 10-200$, the amplitude is of the order of 0.2-0.04. Thus one can expect a 20-4\% deviations coming from finite-size effects.

\begin{figure}[htb!]
\epsscale{1.0}
\plotone{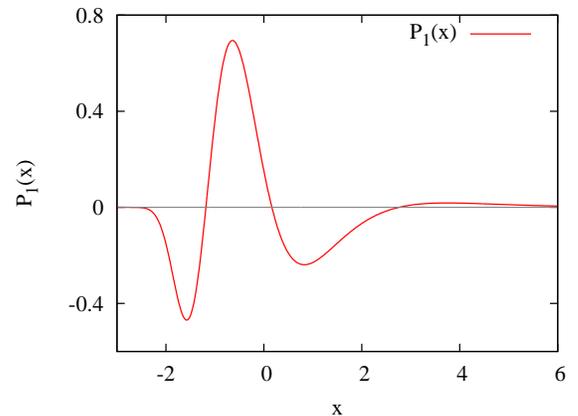}
\caption{First order finite size shape correction function in finite-size scaling of EVS with the Schechter function being the parent distribution. The amplitude of this correction is of the order of $1/\ln N$ for $\beta\neq 1$ while it is of the order of $1/\ln^2 N$ if $\beta=1$.}
\label{FSS-PS}
\end{figure}

The finite-size shape correction is also known \citep{Gyorgyi4}:
\begin{equation}
P_1(x)=\left[P(x)\left[-\frac{ax^2}{2}+\frac{\zeta(3)x}{a^2}+\frac{a}{2}\right]\right]' \,  \label{Eq:Phi_1(x)}
\end{equation}
where $a=\pi/\sqrt{6}$ and $\zeta(z)$ is the Riemann zeta function. The function $P_1(x)$ is plotted on Fig.\ref{FSS-PS} and one can see that the first order correction has well defined signs in various regions of $x$.

A special case arises when the parent distribution is of Gumbel type. In this case, the EVD is also a Gumbel, but with no apparent finite size correction. This is due to the fact that the Gumbel distribution is a fixed point in the renormalization theory formalism used for obtaining the first order corrections \citep{Gyorgyi4,Gyorgyi5}. As a result, the deviations should be caused only by random shot noise from the data points.

\subsubsection{Variable batch size}\label{Sec:VariableBatchSize}

Variable sample size raises basic questions about EVS. In particular, the limiting procedure of sample size going to infinity becomes a problem. If the normalized distribution of $N$ is known ${\cal F}(N)$
then it is natural to consider the average $\overline N =\int {\cal F}(N)NdN$ as the parameter corresponding to the fixed sample size of the usual EVS. Therefore, using the limit $\overline N \to \infty$,
the extreme value distribution becomes
\begin{equation}
\overline P(v)= \lim_{\overline N \to\infty}\int_{0}^\infty {\cal F}(N)P_N(v)dN \, .
\end{equation}
Once $\overline P(v)$ is known, we can write it in scaled variables thus obtaining $\overline P(x)$ and the difference $\overline P_{1}(x) \equiv P(x)-\overline P(x)$ provides us an estimate of corrections
coming from the variable sample size. 

The actual calculation of $\overline P(x)$ assumes that we know ${\cal F}(N)$. A simple form of ${\cal F}(N)$ which fits the observed distribution reasonably well (see Fig.\ref{Number-distr}) and allows analytic calculations is given by
\begin{equation}
{\cal F}(N)=\frac{(N/\overline N-d)^k}{\overline N \Gamma(k+1)}\exp(-N/\overline N + d),
\label{variablesize}
\end{equation}
where $k=3$ and $d$ is a free parameter distinct from zero, since there is a finite cut in $N$ ($N>N_0=\overline{N}$). Note that here we assumed that the distribution can be written in a scaled form 
\begin{equation}
{\cal F}(N)=f(N\,/\,\overline N)/\,\overline N \, .
\end{equation}
This is a good approximation to all of the experimental distributions. Using the above ${\cal F}(N)$, one finds that the limit distribution is universal within the FTG class (and so, for the Schechter function parent distribution as well)
\begin{equation}
\overline{P}(v) = \frac{ \left[  k+1+d(1+e^{-v}) \right] \exp( -de^{-v}-v)  }{ \left[1+e^{-v}\right]^{k+2}  } .
\end{equation}\label{eq:SpecialCase}
The appropriately scaled distribution ($x$ variable) for the case of $d=0$ and $k=3$ is given by the following expression
\begin{equation}
\overline{P}(x)=\frac{4 a \exp(-a x - b)} {(1 + \exp(-a x - b))^5}\, ,
\end{equation}
where $a = \sqrt{ \frac{\pi^2}{3} - \frac{49}{36} }$  and $b = \frac{11}{6}$.

\begin{figure}[htb!]
\epsscale{1.0}
\plotone{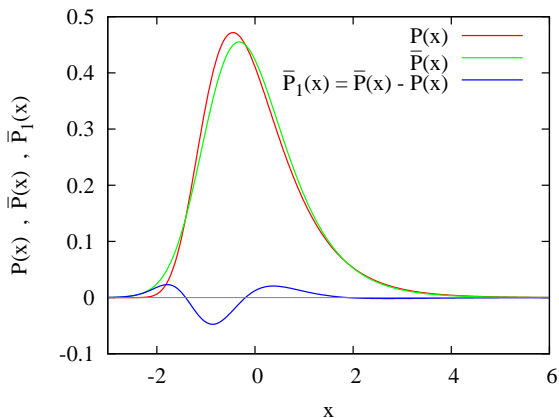}
\caption{Comparison of the FTG limit distribution (red) with that 
obtained from the variable sample size case with the sample-size distribution given by Eq.(\ref{variablesize}) (green line, for the case of $d=0$ and $k=3$). The difference 
of the two functions is also shown, being in the order of 10\%.}
\label{VariN}
\end{figure}

The functions $\overline P(x)$ and $P(x)$ (FTG), and their difference is displayed on Fig.\ref{VariN}. We can see that the maximal difference $\overline P_{1}(x) = \overline P(x)-P(x)$ is of the order of 10\%. What is more interesting is that the positive and negative regions of the differences are significantly shifted compared to those of the finite size corrections (Fig.\ref{FSS-PS}). Thus the two correction may amplify as well as cancel each other, depending on the parent distribution and on ${\cal F}(N)$. 
%\newline
%\newline

%%%%%%%%%%%%%%%%%%%%%%%%%%%%%%%%%%%%%%%%%%%%%%%%%%%%%%%%%%%%%%%%%%%%%%%%%%%%%%%%%%%%%%%%%%%%%%%%%%%%%%%%%%%%%%%%%%%%%%%%%%%%%%%%%%%%%%%%%%%%%%%%%%%%%%%%%%%%%%%%%%%%%%%%%%%%%%%%%%%%
%%%%%%%%%%%%%%%%%%%%%%%%%%%%%%%%%%%%%%%%%%%%%%%%%%%%%%%%%%%%%%%%%%%%%%%%%%%%%%%%%%%%%%%%%%%%%%%%%%%%%%%%%%%%%%%%%%%%%%%%%%%%%%%%%%%%%%%%%%%%%%%%%%%%%%%%%%%%%%%%%%%%%%%%%%%%%%%%%%%%

\section{Distribution of Maximal Luminosities and the Empirical First Order Corrections} \label{Section:DistributionOfMaximalLuminosities}

In order to compute statistics on the maximal luminosities, we used the 2 sampling methods (random sampling and HEALPix-based batches) explained in section \ref{section:FootPrintCreation}.
As we deal with a fixed number $N_{g}$ of galaxies, there is a bias-variance trade-off in all statistics calculated. In fact, increasing the number of batches $n$ does indeed decrease the variance. However, at the same time the data points per batch $N$ decreases, which departs us from the ideal case of $N\rightarrow \infty$, having an increase in the bias. All the statistics are then subject to the balance between $N$ and $n$.

One interesting statistic we measured empirically is the tail index $\xi$ on the extreme value distribution in Eq. \eqref{eq:GEVluminosity}, which is the probability density distribution associated to Eq. \eqref{Eq:CumulativeEVD}. The importance of $\xi$ is that it specifies whether the parent distribution has an infinite reaching tail ($\xi\ge 0$) or a finite cut ($\xi<0 $) at a certain maximum luminosity. 
Next, we measured as well the first order finite size correction for the random sampling method, as well as the influence of the variable batch size in the HEALPix-based method.

\begin{figure*}[htb]
\begin{center}
\includegraphics*[width=8.5cm]{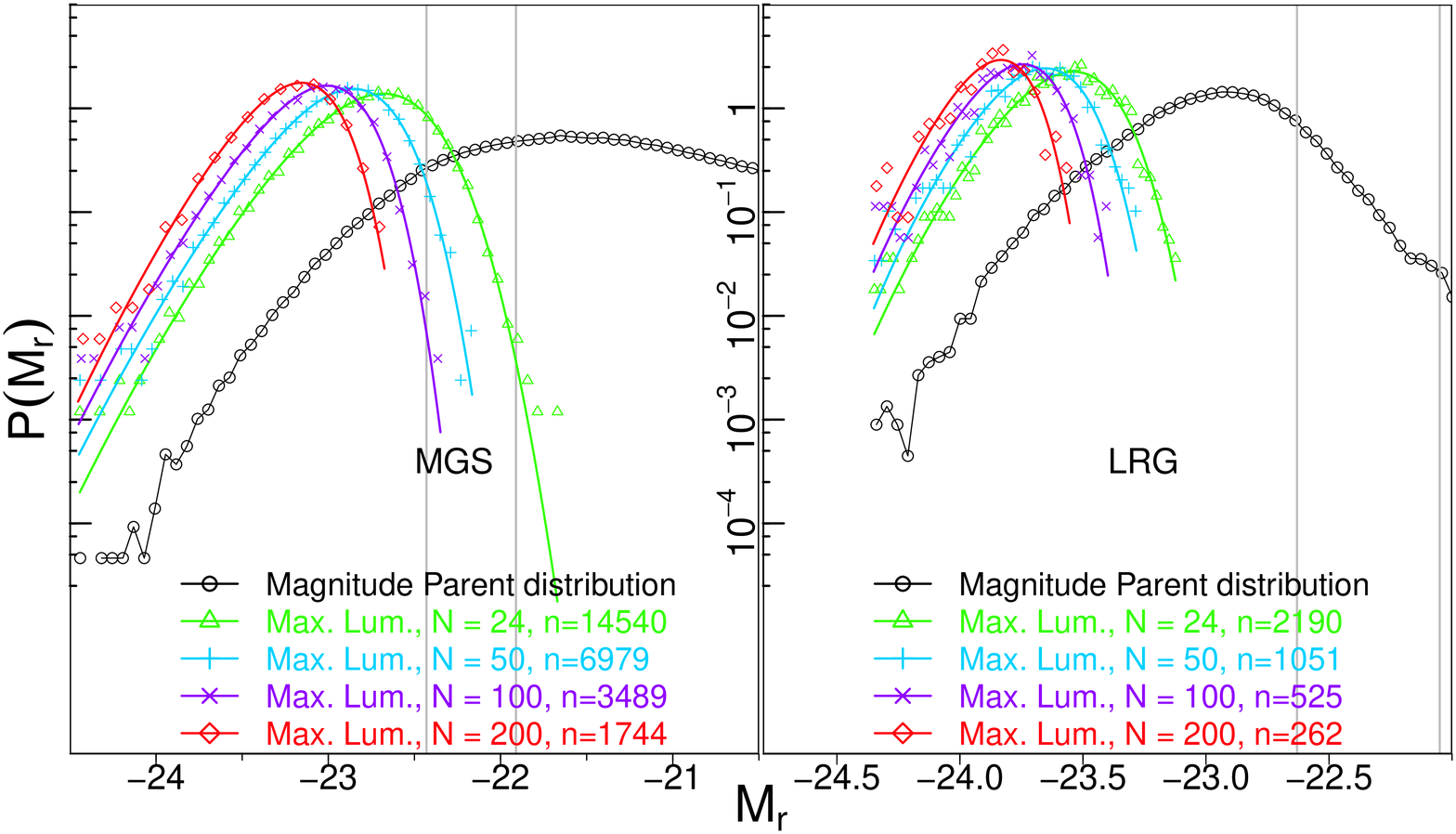}
\includegraphics*[width=7.5cm]{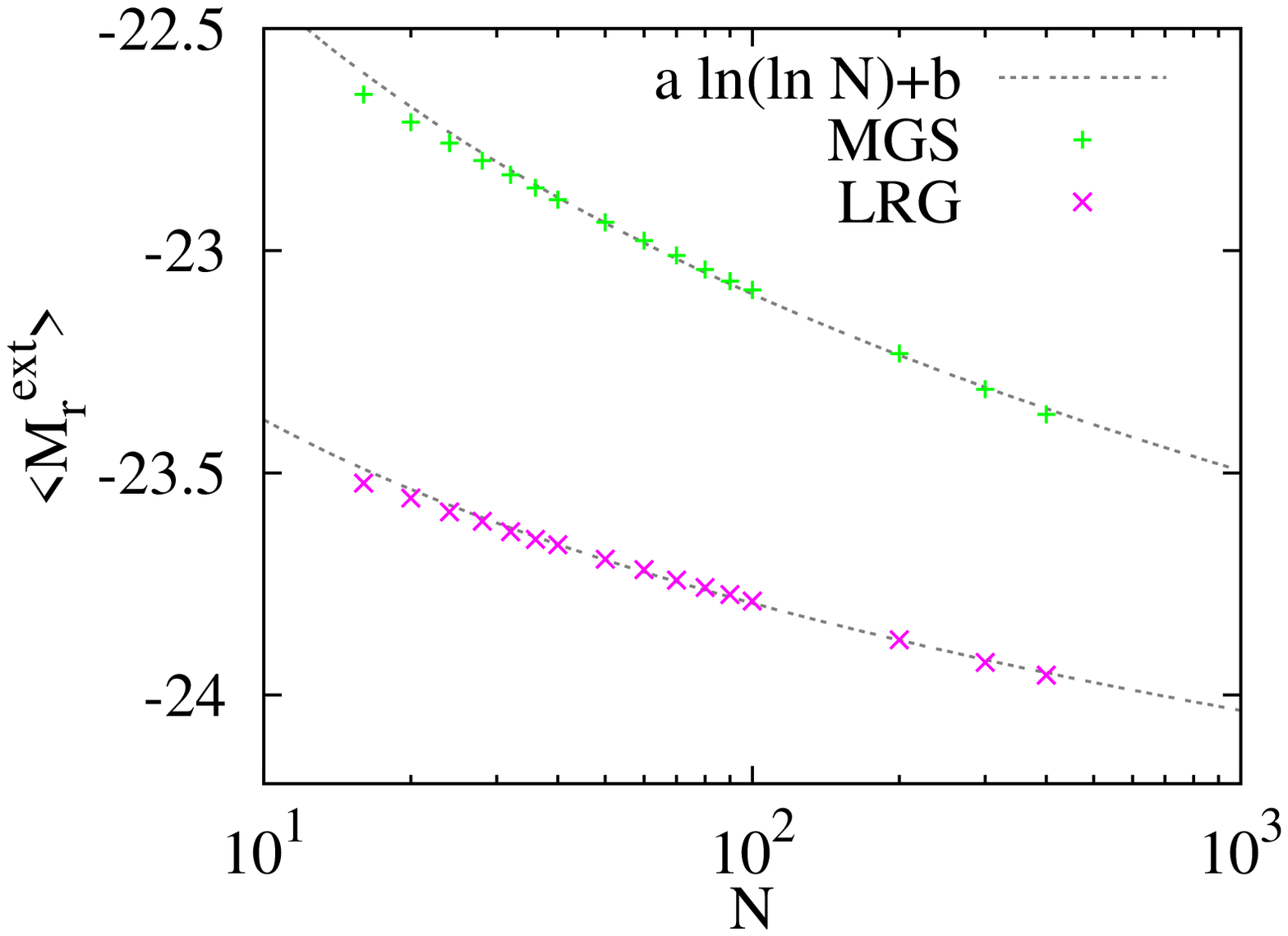}
\end{center}
\caption{$Left$:Distribution of r-band absolute magnitude $M_{r}$ of the full samples. Also included are the distribution of the maximal luminosity for each sample according to the fixed batch size sampling. The vertical lines show the points where $w_{i}=1$ (left line) and $w_{i}=2$ (right line). $Right$: Mean of the maximal luminosity distributions (in absolute magnitude space) v/s batch size $N$ using the fixed batch size sampling. The dashed lines are fits to the asymptotes $\langle M_r^{ext} \rangle  \sim \ln{\ln{N}}$ following from the EVS theory for i.i.d. variables with exponentially decaying parent distributions.}
\label{fig:MagDistribution2}
\end{figure*}

\subsection{Statistics from random sampling batches}\label{sec:ResultsRandomSampling}

\subsubsection{The Tail index $\xi$}\label{sec:TailIndex}

The tail index $\xi$ can be readily calculated in standard EVS using the maximum likelihood estimator on Eq. \eqref{eq:GEVluminosity},i.e., we find numerically the values $L_{\mu}$,$L_{\sigma}$ and $\xi$ that maximize $\ln \Pi_{i=1}^{i=N_{g}} {\rm EVD}(L_{i} | L_{\mu},L_{\sigma},\xi) $ using the Nelder-Mead algorithm \citep{press2007}. We tried this for various combinations of $n$ and $N$. The fitted parameters are in Table \ref{table:FittingParametersEVD}, with probability distributions of the maximal luminosities (in magnitude-space) shown in Figure \ref{fig:MagDistribution2}.
Note that the maximal luminosities are mostly sampled in the region where $w_{i}\leq 2$, which assure us that we are sampling also from the luminosity function.
Note the good overall fit to the EVDs, as well as the increase of $L_{\mu}$ as $N$ increases. The value of $L_{\mu}$ for the LRGs is about twice the size of that for the MGS. As the amount $n$ of batches decreases with $N$, the dispersion and errors in the parameters also increase at higher $N$ as expected. 

For the MGS sample, the value of $\xi$ seems to be positive but very close to zero, with $\xi<0$ being unlikely to happen. Note that for the MGS, the tail index decreases with increasing $N$, so the deviation of $\xi$ from zero may be just a finite size effect. In fact, we can observe that $\xi\sim q(N)$, which is actually the theoretical prediction if we assume a FTG EVD \citep{Gyorgyi5}. The case of the LRG is quite clear, where the value of the tail index does not have a dependence on $N$, having $\xi \approx 0$ within the errors.

Fig.9 also shows the averages of the maximum magnitude
$\langle M^{ext}\rangle$ as function of the batch-size $N$.
As one can see, the results for both the MGS and LRG
samples are well fitted by the theoretical large-$N$
asymptote $a\ln{(\ln{N})}+b$ which follows from the EVS of
an exponential parent distribution for the luminosities.
The test of the theory, however, is not very stringent
since $N$ varies less than 1 and 1/2 decades.

\begin{table}{}
\begin{center}
\caption{Fitting parameters for maximal luminosities\rlap{$^a$}.}
\scriptsize \addtolength{\tabcolsep}{-2pt}
\begin{tabular}{cccccc}
\hline
Sample & $N$	&	$n$	&	$L_{\mu}$							&	$L_{\sigma}$								&	$\xi$	\\
		 &			&			&	${\tiny [10^{10}L_{\odot}]}$	&	${\tiny [10^{10}L_{\odot}]}$			&			\\
\hline
MGS & 24 & 14540 & 7.99 $ \pm $  0.03 & 1.98 $ \pm $ 0.03 & 0.086 $ \pm $ 0.011\\ 
($N_{g}$=348975) & 50 & 6979 & 9.48 $ \pm $  0.05 & 2.10 $ \pm $ 0.04 & 0.091 $ \pm $ 0.017\\ 
 & 100 & 3489 & 10.96 $ \pm $  0.08 & 2.26 $ \pm $ 0.06 & 0.089 $ \pm $ 0.023\\ 
 & 200 & 1744 & 12.59 $ \pm $  0.12 & 2.43 $ \pm $ 0.09 & 0.079 $ \pm $ 0.033\\ 
\hline
LRG & 24 & 2190 & 17.91 $ \pm $  0.12 & 2.72 $ \pm $ 0.09 & 0.002 $ \pm $ 0.028\\ 
($N_{g}$=52579) & 50 & 1051 & 19.86 $ \pm $  0.18 & 2.81 $ \pm $ 0.13 & -0.016 $ \pm $ 0.038\\ 
 & 100 & 525 & 21.75 $ \pm $  0.26 & 2.80 $ \pm $ 0.18 & -0.008 $ \pm $ 0.054\\ 
 & 200 & 262 & 23.69 $ \pm $  0.36 & 2.77 $ \pm $ 0.25 & -0.010 $ \pm $ 0.075\\ 
\hline
\hline
\end{tabular}
\label{table:FittingParametersEVD}
\end{center}
{\small $^a$~ Parameters from the maximum likelihood fitting of the extreme value distribution in Eq. \ref{eq:GEVluminosity}. Maximal luminosity values are sampled at $n$ batches of fixed size $N$. Quoted are the 1-$\sigma$ standard errors. $N_{g}$ denotes the total number of galaxies in each sample.}
\end{table}

\subsubsection{The First order Finite Size Correction}\label{sec:FirstOrderFiniteSizeCorr}

\begin{figure}[h]
\epsscale{1.15}
%\plotone{f11.eps}
\plotone{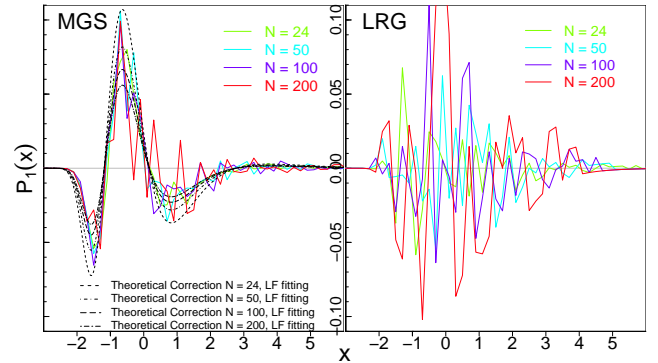}
\caption{Empirical finite size corrections from $n$ batches of fixed size $N$ (from Table \ref{table:FittingParametersEVD}). Each simulation is made by random sampling (without replacement) of a fixed number N of luminosity data points in order to mimic the block maxima approach in EVS. Random sampling with replacement provides similar results. Also, the theoretical first order correction in Eq. \ref{Eq:P_N(x)} is shown for the cases N=24,50,100,200 (with decreasing amplitude).}
\label{Fig:Simulations}
\end{figure}

Motivated by the presence of a finite-$N$, we analyzed the behavior of the empirical finite size corrections for the EVD, and plotted them in Fig. \ref{Fig:Simulations}. Since the estimated values of $\xi$ in the previous section are zero or a small positive number, which is difficult to specify precisely, we assumed for simplicity $\xi=0$ and used the theoretical corrections in \ref{Sec:FiniteSizeCorrection} (theoretical corrections for $\xi \neq 0$ are not developed yet). Here, the empirical corrections are obtained by standardizing the maximal luminosities and subtracting them from the standard Gumbel distribution. 
For plotting the theoretical corrections with an amplitude given in Eq. \eqref{qNforGenGamma}, we need appropriate values of $\alpha$ and $\beta$. As explained in Sec. \ref{Sec:FiniteSizeCorrection}, these fitting parameters should come from fitting the high luminosity tail. Therefore, the fitted values of the full LD in Table \ref{table:FittingParametersLumDistribution} should not be used. In our case, Fig. \ref{fig:LumFun} shows that the full luminosity function fit (Table \ref{table:FittingParametersLumFunction}) is a much better approximation of the high luminosity tail, and we use it instead.

Fig. \ref{Fig:Simulations} shows that the empirical corrections for MGS galaxies do have the same shape as the theoretical first order correction. The amplitude of the function approximately agrees, but we found that the empirical amplitude does not increase significantly when $N$ decreases. The explanation is that as $N$ becomes smaller, we start sampling the maximal luminosities from the bulk of the luminosity distribution instead of its high luminosity tail. Of course, the departure of LF and LD in this regime makes the LF fitting parameters $\alpha$ and $\beta$ no longer valid for calculating the theoretical corrections. A better fit could be attained from $\alpha$ and $\beta$ parameters obtained by fitting the LD to slightly fainted magnitudes than the departing magnitude $M_{D}$. The other consideration is the fact that we might need adding the next term in the correction, which could be important if $N$ is small.

In the LRG case, we cannot find a systematic correction, but just noise. This is in agreement with the expected behavior as explained in Sec. \ref{Sec:FiniteSizeCorrection}, since the Gumbel parent distribution is a fixed point in the renormalization theory used for calculating the corrections \citep{Gyorgyi4,Gyorgyi5}.

%%%%%%%%%%%%%%%%%%%%%%%%

\subsection{Statistics from HEALPix-based batches}\label{sec:ResultsHealPixSampling}

\begin{figure}[h]
\epsscale{1.15}
\plotone{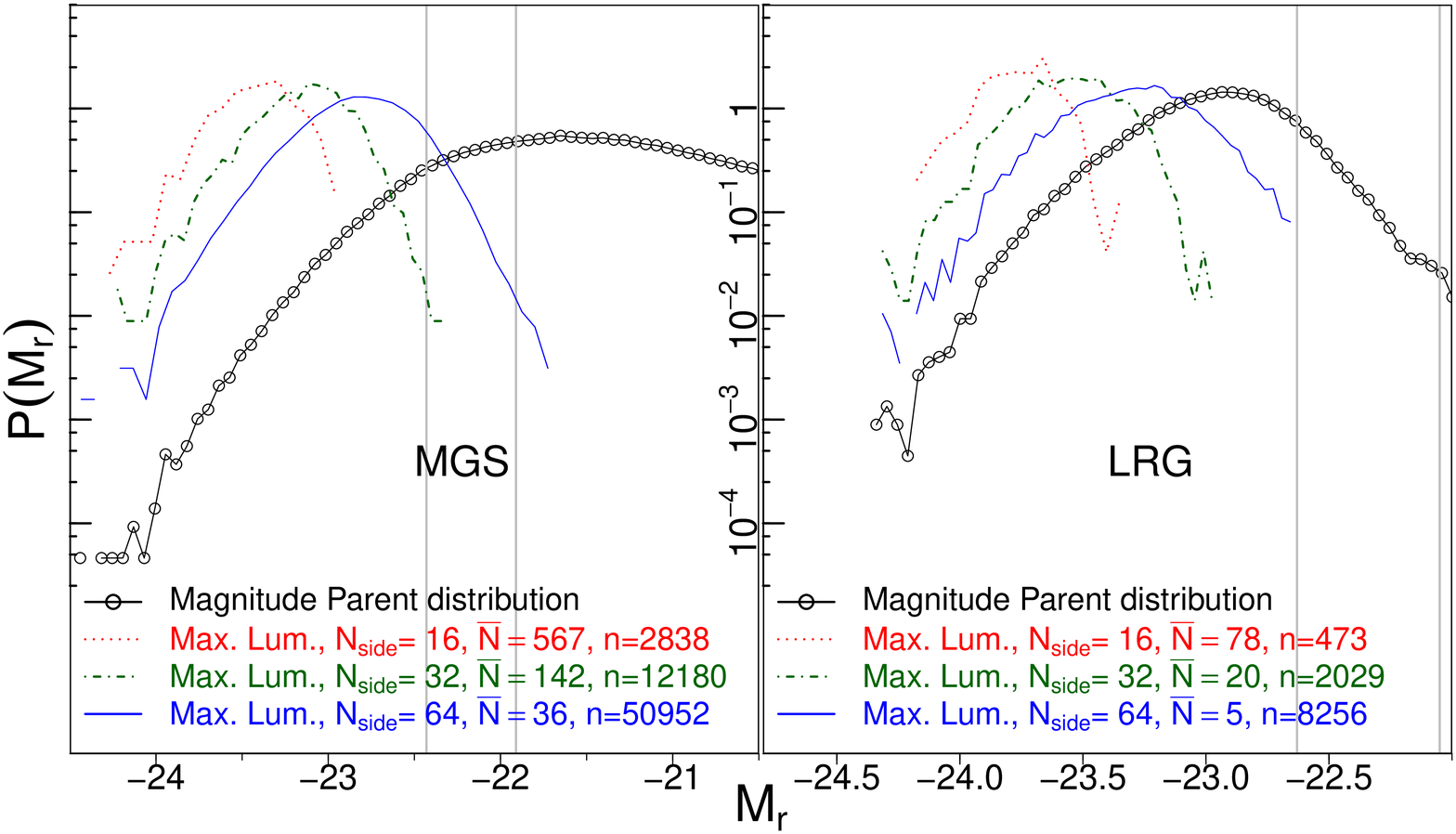}
\caption{Parent distribution of r-band absolute magnitude $M_{r}$ for the full samples. Also included are the distribution of the maximal luminosity for each sample according to different HEALPix resolutions. The vertical lines show the points where $w_{i}=1$ (left line) and $w_{i}=2$ (right line).}
\label{fig:MagDistribution}
\end{figure}

The distributions of the maximal luminosities (in magnitude space) for the HEALPix-based method are shown in Fig. \ref{fig:MagDistribution}. Here, the low luminosity tails of the  maximal distributions reach farther into the low luminosity regions (around $M_{r} \simeq M_{D}$) than in the random sampling method. The reason is that some of the HEALPix cells have very low values of $N$. The maximal luminosities, however, are mostly sampled in the region where $w_{i}\lesssim 2$. As this is the high luminosity region where the LD and LF mostly coincide, all the results obtained from analyzing the bright end of the LD can be also extended and associated with the LF.

\begin{figure}[t]
\begin{center}
\epsscale{1.15}
\plotone{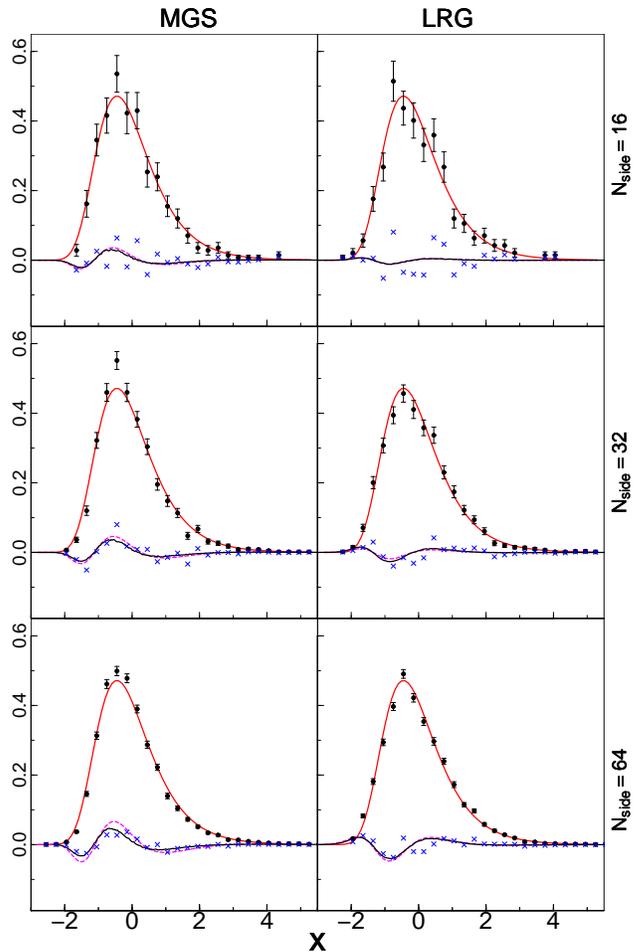}
\end{center}
\caption{The normalized maximum luminosity histograms (black circles) for $N_{side}$ = 16, 32, 64 (from up to down) for the four galaxy samples compared to the limit distribution FTG (solid red line) in scaled variables ($\left\langle x\right\rangle=0$ and $\sigma=1$) while blue crosses are the residuals to the FTG. For the MGS, the solid magenta curves show $q(N)P_{1}(x) + \overline{P}_{1}(x)$, i.e., the first order finite size correction for the Schechter parent added to the variable batch size correction calculated for \eqref{variablesize}.
The LRG curve is different, in the sense that the parent is FTG and the finite size corrections do not appear, having corrections only due to the variable batch size ($\overline{P}_{1}(x)$). 
The black solid curves are the simulations that result from using the experimentally given luminosity distributions and sample size distributions.}
\label{fig:Shape-corrections}
\end{figure}

Fig. \ref{fig:Shape-corrections} presents the distributions of maximal luminosities (in luminosity space) observed in a HEALPix cell for the three studied resolutions ($N_{side}=16,32,64$) and for the two galaxy populations. The distributions on this figure are scaled to zero mean and unite deviation in order to compare them with the similarly scaled Gumbel distribution. The theoretical first order correction coming from the finite-size (finite $N$) effects, together with the correction to the i.i.d. limit distribution coming from the distribution ${\cal F}(N)$ of the galaxy counts are also shown on Fig. \ref{fig:Shape-corrections}. Except for the cases of $N_{side}=16$, where the statistical noise has larger amplitude than the corrections, it appears that the sum of this two corrections is of the order of the residuals and have the same functional shape. At the highest order resolution of the maps ($N_{size}=64$), however, the batch sizes are rather small and the finite-size corrections become large, and appear to be in accord with the theoretical predictions.

The LRGs are special in the sense that finite size effects do not emerge in their EVS, because the parent distribution itself is Gumbel, as explained in \ref{Sec:FiniteSizeCorrection}. Consequently the only correction to the limit distribution comes from the variable batch size $N$, in agreement what we see in the fourth column in Fig. \ref{fig:Shape-corrections}.

We carried out simulations as well in order to confirm our theoretical results by modeling the empirical situation with less noise. Sampling the fitted functions of the empirical parent and sample size distributions with high statistics ($n=10^{7}$) we get smoother histograms. As can be seen on Fig. \ref{fig:Shape-corrections}, the simulated corrections are indeed a good model for the empirical corrections, and support the theoretical expectations as we can observe the convergence toward the theoretical curve for increasing $\overline{N}$.

%%%%%%%%%%%%%%%%%%%%%%%%%%%%%%%%%%%%%%%%%%%%%%%%%%%%%%%%%%%%%%%%%%%%%%%%%%%%%%%%%%%%%%%%%%%%%%%%%%%%%%%%%%%%%%%%%%%%%%%%%%%%%%%%%%%%%%%%%%%%%%%%%%%%%%%%%%%%%%%%%%%%%%%%%%%%%%%%%%%%
%%%%%%%%%%%%%%%%%%%%%%%%%%%%%%%%%%%%%%%%%%%%%%%%%%%%%%%%%%%%%%%%%%%%%%%%%%%%%%%%%%%%%%%%%%%%%%%%%%%%%%%%%%%%%%%%%%%%%%%%%%%%%%%%%%%%%%%%%%%%%%%%%%%%%%%%%%%%%%%%%%%%%%%%%%%%%%%%%%%%

\section{Discussion and Conclusion} \label{disc}

Studying extreme statistics may have several outcomes. One may discover that the objects under consideration have a well defined i.i.d. type extreme value distribution. This may then lead to the conclusion (provided the distribution is of the Weibull type - i.e. the shape parameter is negative) that the underlying objects have an intrinsic cutoff in size. In our case the luminosities have an i.i.d. EVS, but the shape parameter $\xi$ is in the positive range and very close to zero. Thus our conclusion here is that the MGS and LRG luminosities do not have an upper cutoff. 

As far as the LRGs are concerned, we should note that the same conclusion about the absence of an upper cutoff can be reached by a straightforward fit to the high end of the luminosity function. On the other hand, there are difficulties with the agreement of the Schechter fitting to the bright end of the MGS LF \citep[e.g.][]{madgwick2002,bell2003,blanton2003,smith2009}. Here the EVS analysis suggests that the root of the problem may be a small positive tail parameter $\xi$. This possibility was also noted by \cite{alcaniz2004} with their proposal of the generalized double power law fitting function for the LF.

Of course, the conclusion that MGS galaxies do not have a finite luminosity cutoff and $\xi$ has a small positive value is valid only if the methods used in the study are robust against possible corrections arising in the analysis. Uncertainties may come from the finite size of the sample, from the distribution of the number of objects in the sample, and from the correlations among the objects. We have taken care of the finite-size effects by including the first order corrections in the limit distributions and, furthermore, we handled the fluctuations in the sample size by explicitly calculating their effect for the i.i.d. case. 

As evidenced by Figs. \ref{Fig:Simulations} and \ref{fig:Shape-corrections}, a parameter free comparison with the data suggests good agreement with the corrections (for the case $\xi=0$) being the right order of magnitude as well as of the right shape. Thus, we believe that the above effects are in agreement with the conclusion about the absence of upper cutoff in the luminosity. 
Of course, the agreement proved to be valid when the fitting parameters come from a parent distribution fitted in the high luminosity tail, as expected from the theory. If the batch size $N$ decreases and the peak of the maximal luminosities moves into the lower luminosity region, agreement with theory should obtained only if the fitting is performed in an extended luminosity interval considering the lower luminosity values. Since the LF is basically constructed from a weighted LD, the strategy of sampling the maxima from the bright luminosity tail of the LD (where both the LF and LD coincide) was a key part of our analysis. Thus it would be of interest for future studies to develop an extended extreme value theory, where the all data points coming from a given class of parent distributions are each counted with different weights. Such a theory may help in analyzing data sets where there is incompleteness even at the tail where the extremes are sampled from.

The correlations pose a more difficult problem. For one dimensional systems, it is known from the studies of $1/f^\alpha$ type signals that the correlations are irrelevant if they are "weak" \citep{Gyorgyi3}. 
Weak means that the integral of the correlation function is finite. The effectively one-dimensionality of the pencil beam geometry considered in this paper allows the application of the weakness criteria 
for the luminosity correlations. Indeed, one may argue that the luminosity correlations $C_L(r)$ are proportional to the density correlations $C_\rho (r)$ which, at large distances decay as $C_L(r)\sim C_\rho (r) \sim 1/r^2$. The one-dimensional integral of this type of correlations is convergent, thus we believe the weakness criteria is satisfied, and our conclusion is not affected by the correlations
[note that any power relationship between the correlations ($C_L(r)\sim C_\rho^\mu (r)$) will also satisfy the criteria of weakness provided $\mu >1/2$].

We can thus conclude that the extreme value statistics of galaxy luminosities is i.i.d. type with zero or small positive shape parameter, and this conclusion takes into account the finite-size of the samples, the galaxy-number fluctuations in the pencil beams, and the large-distance spatial correlations among luminosities.

%%%%%%%%%%%%%%%%%%%%%%%%%%%%%%%%%%%%%%%%%%%%%%%%%%%%%%%%%%%%%%%%%%%%%%%%%%%%%%%%%%%%%%%%%%%%%%%%%%%%%%%%%%%%%%%%%%%%%%%%%%%%%%%%%%%%%%%%%%%%%%%%%%%%%%%%%%%%%%%%%%%%%%%%%%%%%%%%%%%%
%%%%%%%%%%%%%%%%%%%%%%%%%%%%%%%%%%%%%%%%%%%%%%%%%%%%%%%%%%%%%%%%%%%%%%%%%%%%%%%%%%%%%%%%%%%%%%%%%%%%%%%%%%%%%%%%%%%%%%%%%%%%%%%%%%%%%%%%%%%%%%%%%%%%%%%%%%%%%%%%%%%%%%%%%%%%%%%%%%%%

\acknowledgments

K. O. and Z. R. have been partially supported by the Hungarian Science Foundation OTKA through grants No. K 68109 and NK100296.
M.T.P thanks Sebastien Heinis, Ching-Wa Yip and Mark Neyrinck for useful discussion.

Funding for SDSS-III has been provided by the Alfred P. Sloan Foundation, the Participating Institutions, the National Science Foundation, and the U.S. Department of Energy Office of Science. The SDSS-III web site is http://www.sdss3.org/.

SDSS-III is managed by the Astrophysical Research Consortium for the Participating Institutions of the SDSS-III Collaboration including the University of Arizona, the Brazilian Participation Group, Brookhaven National Laboratory, University of Cambridge, Carnegie Mellon University, University of Florida, the French Participation Group, the German Participation Group, Harvard University, the Instituto de Astrofisica de Canarias, the Michigan State/Notre Dame/JINA Participation Group, Johns Hopkins University, Lawrence Berkeley National Laboratory, Max Planck Institute for Astrophysics, New Mexico State University, New York University, Ohio State University, Pennsylvania State University, University of Portsmouth, Princeton University, the Spanish Participation Group, University of Tokyo, University of Utah, Vanderbilt University, University of Virginia, University of Washington, and Yale University.


\begin{thebibliography}{77}









\bibitem[Aihara et al.(2011)]{aihara2011} Aihara, H., et al., 2009, \apjs, 193, 29A
\bibitem[Alcaniz \& Lima(2004)]{alcaniz2004} Alcaniz, J. S. \& Lima, J.A.S. 2004, Brazilian Journal of Physics, 34, 2A
\bibitem[Baldry et al.(2004)]{baldry2004} Baldry, I. K., et al. 2004, ApJ, 600, 681
\bibitem[Bell et al.(2003)]{bell2003} Bell, E.~F., McIntosh, D.~H., Katz, N., \& Weinberg, M.~D.\ 2003, \apjs, 149, 289 
\bibitem[Bermann(1964)]{Berman} Berman S. M., Ann. Math. Stat. 33, 502 (1964).

\bibitem[Bernardi et al.(2010)]{bernardi2010} Bernardi,M., Shankar, F., Hyde, J. B., et al. 2010, MNRAS. 404, 2087


\bibitem[Bhavsar \& Barrow(1985)]{bhavsar1985} Bhavsar, S. P. \& Barrow, J. D. 1985, MNRAS, 213, 857
\bibitem[Binggeli, Sandage \& Tammann(1988)]{binggeli1988} Binggeli, B., Sandage, A. \& Tammann, G.A. 1988, ARA\&A, 26, 509 
\bibitem[Blanton et al.(2001)]{blanton2001} Blanton, M. R., Dalcanton, J., Eisenstein, D., et al. 2001, AJ, 121, 2358
\bibitem[Blanton et al.(2003)]{blanton2003} Blanton, M. R., Hogg, D. W., Bahcall, N. A., et al. 2003, ApJ, 592, 819
\bibitem[Bruzual \& Charlot(2003)]{bruzual2003} Bruzual A.,G. \& Charlot ,S., 2003, MNRAS 344, 1000
\bibitem[Budav\'{a}ri et al.(2000)]{budavari2000} Budav\'{a}ri, T., et al., 2000, \aj, 120, 1588
\bibitem[Csabai et al.(2000)]{csabai2000} Csabai, I., Connolly, A. J., Szalay, A.S., \& Budav\'{a}ri, T. 2000, AJ, 119,69
\bibitem[Coles(2001)]{coles2001} Coles, S., (2001), An Introduction to Statistical Modeling of Extreme Values, Springers
\bibitem[Cooray \& Milosavljevic(2005)]{cooray2005} Cooray, A. \& Milosavljevic, M. 2005, ApJ, 627, L89
\bibitem[Cooray(2006)]{cooray2006} Cooray, A. 2006, MNRAS, 365, 842

\bibitem[Croton et al.(2007)]{croton2007} Croton,D., et al. 2007, MNRAS. 379, 1562


\bibitem[De Lucia \& Blaizot(2007)]{delucia2007} De Lucia G. \& Blaizot J. 2007, MNRAS, 375, 2
\bibitem[Dobos \& Csabai(2011)]{dobos2011} Dobos, L. \& Csabai, I. 2011, \mnras, 414, 1862D
\bibitem[Eisenstein et al.(2001)]{eisenstein2001} Eisenstein, D., et al. 2001, ApJ, 122, 2267
\bibitem[Embrechts et al.(1997)]{embrechts1997} Embrechts, P., Kl\"{u}ppelberg, C. \& Mikosch, T. (1997), Modeling extremal events for insurance and finance, Springer, Berlin.
\bibitem[Fioc \& Rocca-Volmerange(1997)]{fioc1997} Fioc, M., \& Rocca-Volmerange, B. 1997, A\&A, 326, 950
\bibitem[Galambos(1978)]{galambos} Galambos, J. 1978, {The Asymptotic Theory of Extreme Order Statistics} (New York: Wiley)
\bibitem[Geller \& Tremaine(1976)]{geller1976} Geller, M. J. \& Peebles, P. J. E. 1976, AJ, 206, 939
\bibitem[G\'{o}rski et al.(2005)]{gorski2005} G\'{o}rski, K. M., et al. 2005, ApJ, 622, 759
\bibitem[Gumbel(1958)]{gumbel} Gumbel, E. 1958, {Statistics of Extremes} (New York: Dover).
\bibitem[Gy\"orgyi et al.(2007)]{Gyorgyi3} Gy\"orgyi, G., Moloney, N. R., Ozog\'any, K. and R\'acz, Z. 2007, Phys. Rev. E 75, 021123
\bibitem[Gy\"orgyi et al.(2008)]{Gyorgyi4} Gy\"orgyi, G., Moloney, N. R., Ozog\'any, K. and R\'acz, Z. 2008, Phys. Rev. Lett. 100, 210601 
\bibitem[Gy\"orgyi et al.(2010)]{Gyorgyi5} Gy\"orgyi, G., Moloney, N. R., Ozog\'any, K., R\'acz, Z. and Droz, M. 2008, Phys. Rev. E , 81, 041135
\bibitem[Lin et al.(2010)]{lin2010} Lin, Y. et al. 2010, \apj, 715, 1486
\bibitem[Loh(2004)]{loh2004} Loh, Yeong-Shang 2004, {Luminous Red Galaxies in the Sloan Digital Sky Survey}, Princeton Physics PhD thesis. 
\bibitem[Loh \& Strauss(2006)]{loh2006} Loh, Yeong-Shang \& Strauss, M. A. 2006, \mnras 366, 373
\bibitem[Madgwick et al.(2002)]{madgwick2002} Madgwick, D. S. et al. 2002, \mnras, 333, 133
\bibitem[Ostriker \& Hausman(1977)]{ostriker1977} Ostriker J. P. \& Hausman M. A., 1977, ApJL, 217, L125
\bibitem[Paranjape \& Sheth(2011)]{paranjape2011} Paranjape, A. \& Sheth, R. K. 2011, arXiv:1107.3652
\bibitem[Postman \& Lauer(1995)]{postman1995} Postman, M. \& Lauer, T. R. 1995, ApJ, 440, 28
\bibitem[Press \& Schechter(1974)]{pressschechter1974} Press,W. H. \& Schechter, P. 1974, \apj, 187,425

\bibitem[Press et al.(2007)]{press2007} Press, W. H, Teukolsky, S. A., Vetterling, W. T \& Flannery, B. P. 2007, Numerical Recipes: The Art of Scientific Computing, Cambridge University Press, New York 

\bibitem[Reiss \& Thomas(1997)]{reiss1997} Reiss, R. \& Thomas, M. (1997), Statistical Analysis of Extreme Values with Applications
to Insurance, Finance, Hydrology and Other Fields, Birkh\"{a}user, Basel.
\bibitem[Schechter(1976)]{schechter1976} Schechter, P. 1976, \apj, 203,297
\bibitem[Schmidt(1968)]{schmidt1968} Schmidt, M. 1968, ApJ, 151, 393 
\bibitem[Shimasaku et al.(2001)]{shimazaku2001} Shimazaku, K., et al. 2001, \aj, 122, 1238
\bibitem[Smith et al.(2009)]{smith2009} Smith, A. J., et al. 2009, \mnras, 397, 868
\bibitem[Stoughton et al.(2002)]{stoughton2002} Stoughton, C., Lupton, R.~H., Bernardi, M., et al.\ 2002, \aj, 123, 485
\bibitem[Strateva et al.(2001)]{strateva2001} Strateva, I., et al., 2001, \aj, 122, 1861
\bibitem[{{Strauss} {et~al.}(2002)}]{strauss2002}{Strauss}, M.~A., {et~al.} 2002, \aj, 124, 1810
\bibitem[Taghizadeh-Popp(2010)]{taghizadeh-popp2010} Taghizadeh-Popp, M. 2010, PASP, 122, 976
\bibitem[Tremaine \& Richstone(1977)]{tremaine1977} Tremain, S. D. \& Richstone, D. O. 1977, AJ, 212, 311

\bibitem[{Yang \& Saslaw(2011)}]{yang2011} Yang, A. \& Saslaw, W. C. 2011, \apj, 729, 123


\bibitem[{{York} {et~al.}(2000)}]{york00}{York}, D.~G., {et~al.} 2000, \aj, 120, 1579




\end{thebibliography}
\end{document}